\documentclass[%
 reprint,
 nofootinbib,
 amsmath,amssymb,
 aps,
]{revtex4-2}
\pdfoutput=1

\usepackage{graphicx}
\usepackage{dcolumn}
\usepackage{bm}
\usepackage[caption=false]{subfig}
\usepackage{xcolor}
\usepackage{newtxtext,newtxmath}
\usepackage[T1]{fontenc}
\usepackage{tabularx}
\usepackage{cancel}
\usepackage{aas_macros}
\usepackage{caption}
\usepackage{array}

\begin{document}

\title{The REACH radiometer for detecting the 21-cm hydrogen signal from redshift $\sim$ 7.5 to 28}

\author{E. de Lera Acedo,$^{1,2}$
D.I.L. de Villiers,$^{3}$
N. Razavi-Ghods,$^{1}$
W. Handley,$^{1,2}$
A. Fialkov,$^{2,4}$
A. Magro,$^{5}$
D. Anstey,$^{1}$\\
H.T.J. Bevins,$^{1}$
R. Chiello,$^{6}$
J. Cumner,$^{1}$
A.T. Josaitis,$^{1}$
I.L.V. Roque,$^{1}$
P.H. Sims,$^{7,8}$
K.H. Scheutwinkel,$^{1}$\\
P. Alexander,$^{1}$
G. Bernardi,$^{9,10,11}$
S. Carey,$^{1}$
J. Cavillot,$^{12}$
W. Croukamp,$^{3}$
J.A. Ely,$^{1}$
T. Gessey-Jones,$^{1}$\\
Q. Gueuning,$^{1}$
R. Hills,$^{1}$
G. Kulkarni,$^{13}$
R. Maiolino,$^{1,2}$
P. D.  Meerburg,$^{14}$
S. Mittal,$^{13}$
J.R. Pritchard,$^{15}$\\
E. Puchwein,$^{16}$
A. Saxena,$^{14}$
E. Shen,$^{1}$
O. Smirnov,$^{10,11}$
M. Spinelli,$^{17,18,19}$
K. Zarb-Adami,$^{5,6}$\\
\hfill \break
$^{1}$Cavendish Astrophysics, University of Cambridge, Cambridge, UK\\
$^{2}$Kavli Institute for Cosmology in Cambridge, University of Cambridge, Cambridge, UK\\
$^{3}$Department of Electrical and Electronic Engineering, Stellenbosch University, Stellenbosch, South Africa\\
$^{4}$Institute of Astronomy, University of Cambridge, Cambridge, UK\\
$^{5}$Institute of Space Sciences and Astronomy, University of Malta, Malta\\
$^{6}$Physics Department, University of Oxford, Oxford, UK\\
$^{7}$McGill Space Institute, McGill University, Montr\'eal, Canada\\
$^{8}$Department of Physics, McGill University, Montr\'eal, Canada\\
$^{9}$INAF-Istituto di Radio Astronomia, via Gobetti 101, 40129 Bologna, Italy\\
$^{10}$Department of Physics and Electronics, Rhodes University, PO Box 94, Grahamstown, 6140, South Africa\\
$^{11}$South African Radio Astronomy Observatory, Black River Park, 2 Fir Street, Observatory, Cape Town, 7925, South Africa\\
$^{12}$Antenna Group, Université catholique de Louvain, Belgium\\
$^{13}$Department of Theoretical Physics, Tata Institute of Fundamental Research, Mumbai, India\\
$^{14}$Faculty of Science and Engineering, University of Groningen, Groningen, The Netherlands\\
$^{15}$Department of Physics, Imperial College London, London, UK\\
$^{16}$Leibniz Institute for Astrophysics, Potsdam, Germany\\
$^{17}$INAF-Osservatorio Astronomico di Trieste, Via G.B. Tiepolo 11, I-34143 Trieste, Italy\\
$^{18}$ IFPU - Institute for Fundamental Physics of the Universe, Via Beirut 2, 34014 Trieste, Italy\\
$^{19}$ Department of Physics and Astronomy, University of the Western Cape, Robert Sobukhwe Road, Bellville, 7535, South Africa\\
\hfill \break
}

\collaboration{REACH Collaboration}


\begin{abstract}
Observations of the 21-cm line from primordial hydrogen promise to be one of the best tools to study the early epochs of the Universe: the Dark Ages, the Cosmic Dawn, and the subsequent Epoch of Reionization. In 2018, the EDGES experiment caught the attention of the cosmology community with a potential detection of an absorption feature in the sky-averaged radio spectrum centred at 78 MHz. The feature is deeper than expected, and, if confirmed, would call for new physics. However, different groups have re-analyzed the EDGES data and questioned the reliability of the signal. The Radio Experiment for the Analysis of Cosmic Hydrogen (REACH) is a sky-averaged 21-cm experiment aiming at improving the current observations by tackling the issues faced by current instruments related to residual systematic signals in the data. The novel experimental approach focuses on detecting and jointly explaining these systematics together with the foregrounds and the cosmological signal using Bayesian statistics. To achieve this, REACH features simultaneous observations with two different antennas, an ultra wideband system (redshift range $\sim$ 7.5 to 28), and a receiver calibrator based on in-field measurements. Simulated observations forecast percent-level constraints on astrophysical parameters, potentially opening up a new window to the infant Universe.

\end{abstract}

\maketitle

\section{\label{sec:intro}Introduction}

\begin{figure*}[!htb]
\centerline{\includegraphics[width=17cm]{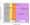}}
\caption{\label{fig:21-cmline}A typical model of the Global 21-cm line with highlights of the main cosmic events \citep{furlanetto06b, pritchard06}. From left to right: collisional coupling (grey), onset of Ly-$\alpha$ coupling (yellow), onset of X-ray heating (orange), photo-ionization (purple). REACH will explore the frequency range 50 - 170 MHz ($z$ $\sim$ 7.5-28).}
\end{figure*}

The first stars and galaxies formed some time after the epoch of recombination, when the Cosmic Microwave Background (CMB) decoupled $\sim$378,000 years after the Big Bang at a redshift ($z$) of $\sim$1100, and before the current `realm of galaxies' that we can see today \citep[][]{Naoz:2006, Loeb:2013, Klessen:2019}. The radiation from these first luminous sources heated and reionized the neutral hydrogen that pervaded the primordial Cosmos \citep[][]{Loeb:2013, Barkana:2016}. Probing the intermediate epochs, the ‘Dark Ages’ before the first stars, through first new light in the Universe, and cosmic reionization, represents a frontier in the study of cosmic structure formation~\citep[][]{Furlanetto:2006, Mesinger:2019}. The hyperfine transition of neutral hydrogen has a restframe wavelength of 21 cm and by observing at low radio frequencies we can study directly its redshifted radio emission (and absorption) from the gas clouds that were the raw material that formed the first luminous cosmic structures at these early epochs \citep[][]{Furlanetto:2006, Barkana:2016,  Mesinger:2019}. Figure \ref{fig:21-cmline} shows an example of the sky-averaged (global) 21-cm signal for an assumed succession of cosmic events. However, as high-redshift models are poorly constrained by existing data, a large variety of 21-cm signals are still plausible and the timing, and order, of cosmic events is not well understood \citep[][]{cohen:2017, fialkov18, fialkov19, reis20}.

A number of experiments to measure the global redshifted 21-cm signal are already underway, including EDGES \citep[Experiment to Detect the Global Epoch of Reionization Signature,][]{2008ApJ...676....1B}, SARAS \citep[Shaped Antenna measurement of the background RAdio Spectrum,][]{2018ApJ...858...54S}, PRIZM~\citep[Probing Radio Intensity at High-Z from Marion,][]{Philip2019}, LEDA \citep[Large-aperture Experiment to Detect the Dark Ages,][]{Bernardi2016}, SCI-H I \citep[Sonda Cosmológica de las Islas para la Detección de Hidrógeno Neutro,][]{2014ApJ...782L...9V}, or have reported results in the recent years, such as BIGHORNS \citep[Broadband Instrument for Global HydrOgen ReioNisation Signal,][]{2015PASA...32....4S}, or are being planned, like DAPPER (Dark Ages Polarimeter PathfindER, \url{https://www.colorado.edu/project/dark-ages-polarimeter-pathfinder/}) and MIST (Mapper of the IGM Spin Temperature, \url{http://www.physics.mcgill.ca/mist/}) among others. The deep  absorption profile recently detected by the EDGES collaboration at a frequency of 78 MHz  \citep{bowman18} is the first candidate detection of the Global 21-cm signal from  $z\sim 17$. If truly of cosmological origin as opposed to being caused by instrumental systematical errors or by foreground signals, this signal implies significant star formation at high redshifts \citep{fialkov19, mirocha19, Schauer:2019, reis20} and, thus, could be the first real evidence of the primordial population of stars. However, the amplitude of the detected signal of  $500^{+500}_{-200}$ mK (at 99\% confidence) is too strong compared to the standard expectations obtained with conventional astrophysical modelling \citep{cohen:2017,cohen19, Reis2021}. The standard scenario \citep{furlanetto06b, pritchard06} assumes that the atomic hydrogen gas is, first, cooled down by the expansion of the Universe to temperatures well below that of the CMB and, then, is heated by X-ray sources. Lyman-alpha (Ly-$\alpha$) photons produced by the first population of stars couple the 21-cm spin temperature to the  kinetic temperature of the gas, thus rendering the 21-cm signal visible against the background radiation via the process called Wouthuysen-Field (WF) coupling \citep{wouthuysen52, field58}.  It is standard practice to assume that the 21-cm signal is observed against the CMB, and in this picture the depth of the absorption trough is at most $\sim200$ mK at $z\sim17$~\citep{fialkov19}, with the deepest features achieved assuming saturated WF coupling and gas cooled by the adiabatic expansion of the Universe in the absence of X-ray (or other) heating sources. If not explained by hardware systematics, the anomalously deep feature reported by EDGES calls for exotic theoretical interpretations (see more in section \textit{Methods / Exotic interpretations of the EDGES signal}).

The REACH experiment, introduced in this work, was proposed  to independently measure the high-redshift 21-cm signal in the $\nu = 50-170$ MHz frequency range ($z$ $\sim$ 7.5-28), and, thus, either verify or disprove the EDGES detection. The REACH pipeline will include a variety of theoretical models (both standard and exotic ones). If a cosmological 21-cm  signal is detected, these models will allow us to measure/constrain astrophysical parameters associated with primordial star and black hole formation, thermal and ionization histories of the Universe{\color{red},} and radio production efficiency of high-redshift astrophysical sources. Specifically, the location and depth of the absorption trough in the 21-cm signal can tell us (i) the timing of primordial star formation, (ii) the properties of the first star forming objects (mass and star formation efficiency), and (iii) the luminosity of the first X-ray sources (e.g., the first population of X-ray binaries). REACH will also be able to tell us whether the standard astrophysical picture has to be revised by, for instance,  adding an excess radio background and/or extra  cooling at high redshifts. In the case of a non-detection, the upper limits on the strength of the absorption feature within the REACH frequency band can be used to bound high-redshift astrophysics. Low-intensity signals within the REACH band would require either the X-ray sources to be very luminous or for star formation to happen very late and in very massive dark matter halos. In addition, models that assume extreme radio background and/or dark matter over-cooling would be considered less likely.

Re-analysis of public EDGES data has resulted in concerns about the data analysis presented by EDGES, which uses a fit with a basic foreground model with non-physical parameters \citep{2018Natur.564E..32H}. Recent studies \citep{2019ApJ...880...26S, 2020MNRAS.492...22S, Bevins2020} have also revealed alternative ways of explaining the data in terms of residual instrumental systematics as shown in Extended Data Fig. 1 and described with more detail in section \textit{Methods / EDGES data re-analysis}. Recently, the SARAS collaboration has published an analysis of data collected with their latest antenna design (a monocone antenna floating on lake water) showing residuals incompatible with the cosmological signal profile reported by EDGES \citep{singh2021detection}. 

In order to solve this puzzle and aim at a confident detection of the 21-cm signal, REACH has been designed to avoid systematic signals potentially degenerate with the cosmological signal and to detect any residual systematic signals and model them in the data together with the cosmological fits. This is achieved by using physically-based foreground and instrument models that will be jointly fitted, using Bayesian inference, with the cosmic signal models. The aim of this approach is to be able to explain any residual instrument systematics and their correlation with the foregrounds and cosmological signals. In the presence of unaccounted systematics, techniques have been developed using Maximally Smooth Functions for their identification and mathematical characterisation. This is in contrast with the first generation sky-averaged experiments (e.g. \citep{2008ApJ...676....1B, 2018ApJ...858...54S}) where the main focus was on developing and operating an instrument that was as achromatic (spectrally smooth) as possible. A summary table comparing the main experimental features of existing experiments with REACH is shown in Tab. \ref{tab:Comparison}.

REACH is a wide band experiment covering both the Cosmic Dawn and the Epoch of Reionization. REACH will use a nested sampling tool, \texttt{PolyChord} \citep{2015MNRAS.453.4384H,2015MNRAS.450L..61H}, and parametric foreground, instrument and 21-cm signal models \citep{cohen:2017} for the signal detection. REACH also features a switched calibrator Radio Frequency (RF) receiver using in-field measurements of the analogue and digital components on the receiving chain. The calibration of the receiver uses a fast Bayesian conjugate prior based approach. REACH Phase I is currently being deployed in the RFI-quiet Karoo radio reserve in South Africa, which is the location of the HERA, MeerKat, and the future SKA1-Mid experiments~\citep{Dewdney_SKA,Jonas_MeerKAT,DeBoer_2017}. REACH Phase I will observe the sky with two independent and different antennas simultaneously. The antenna designs - a conical log-spiral antenna and a hexagonal dipole - have been selected using a simulated pipeline and maximising the Bayesian log evidence for an assumed 21-cm signal model in the received data while minimising the error in the recovered 21-cm signal. More details of this are provided in the~\textit{Methods} section. This paper introduces the experiment; the methodology and experimental approach as well as a brief description of its hardware and software (algorithm) components. As the project evolves and we analyze real sky data and laboratory measurements we expect to validate the approach described here as well as to find that modifications of the different components may be required. A series of papers authored by members of the REACH team describing different components of the experiment in more detail are publicly available. These are referred to throughout the text and include: the description of the Bayesian data analysis pipeline \citep{anstey20}, the description of the Bayesian reciver calibration \citep{reachBayes}, the antenna selection using the Bayesian data pipeline \citep{anstey2021informing}, the design of the Phase I wideband dipole \citep{cumner2021radio}, the assessment of ionospheric effects \citep{Shen2020}, the detection of systematic signals using Maximally Smooth Functions \citep{Bevins2020}, the sky-averaged 21-cm signal emulator \citep{bevins2021globalemu}, the Bayesian evidence-driven likelihood selection \citep[][]{ScheutwinkelA}, and the Bayesian evidence-driven diagnosis of instrumental systematics \citep[][]{ScheutwinkelB}.

REACH is planned to be staged in phases. A Phase II will follow featuring further antenna systems, which could include scaled versions of the Phase I antennas or complementary antenna systems sensitive to other polarizations of the sky radiation, amongst others.

In section~\textit{Experimental Approach} we highlight and briefly discuss the main features of the REACH experiment. Section~\textit{System Design} depicts the high-level hardware design, later extend in the \textit{Methods} section. Section~\textit{Data Analysis and Science Predictions} describes the joint Bayesian data and science analysis pipeline and shows the predicted performance of REACH and its ability to constrain the relevant cosmological models based on a simulated data analysis. We conclude with some final remarks in the~\textit{Conclusion} section and in the~\textit{Methods} section the main methods developed for REACH are explained. 

\section{Experimental approach: Explain the systematics}

The novel experimental approach of REACH is focused on understanding and jointly constraining telescope systematics with the cosmological and foreground signal.

The spectral and spatial structure of the foregrounds couple with the spectral and spatial variations of the antenna on the ground, resulting (even for a simple dipole antenna) in antenna temperature variations which cannot be modelled with a simple low order polynomial and which are highly dependent on Local Sidereal Time (LST) and integration times. Reflections and spectral features in the analogue receiver can also have scales similar to those of the target cosmological signal. If any such systematic feature remains in the data after calibration, it may not be possible to accurately constrain the cosmological models.

In order to achieve a convincing and confident detection of the 21-cm signal we argue that the foreground models should be independent from the instrument models and the cosmological models, and the correlation between their parameters needs to be clearly explained and isolated in the data analysis. This should also be done using physics rooted models, field measurements and supported by the robust statistical inference. 

\textbf{In order to achieve this, REACH uses a Bayesian calibration and data analysis pipeline where correlation between all the different parameters can be explained in a statistical manner.} REACH has been designed to avoid instrumental systematic signals in the first place, but then in order to deal with any residual systematic effects, we have used a simulated version of the data pipeline to design an instrument where all the 3 signal spaces (cosmological signal, foregrounds and instrument) are largely orthogonal \citep[][]{anstey2021informing}. While this approach does not ensure success in achieving the very challenging detection, it provides a framework to enhance the robustness of the data analysis against the effect of instrumental systematic signals. This process however requires a substantial amount of information to constrain the models. In order to provide this extra information, in REACH Phase I we feature the following novelties, unique to this type of experiment:

\begin{itemize}
    \item \textbf{Simultaneous observations with two different antennas.} REACH will observe the southern hemisphere sky with two different radio antennas simultaneously. The data collected by the two antennas will then be analyzed jointly in order to better understand and isolate signal components associated with the hardware systematics. The benefits of doing so are discussed in more detail in section \textit{Methods / Time- and antenna-dependant modelling}. These antennas are, for Phase I, a hexagonal dipole and a conical log spiral antenna. In section \textit{Antennas} we describe the design process for choosing antenna types and optimizing their performance. For Phase II we then expect to deploy further antennas, including scaled versions of the dipole and a dual polarization system. Furthermore, the use of two antennas, even if primarily used to analyze auto-correlated data, will allow us to cross correlate (in an interferometric sense) their output voltages to further constrain hardware systematics.
    \item \textbf{Ultra wideband system.} Separating the foregrounds from the cosmological 21-cm signal relies almost entirely on their different spectral components. The foregrounds, dominated by synchrotron radiation from  our own galaxy, are expected to be smooth \citep{SathyanarayanaRao2017}, well described by a power law in frequency. The sky-averaged 21-cm signal however, as shown in Fig. \ref{fig:21-cmline}, is expected to oscillate in frequency and exhibit three turning points between 50 MHz and 170 MHz. Thus, a larger frequency bandwidth should provide more chances to leverage these spectral differences. First generation 21-cm global experiments \citep{Rogers2008, Singh2018a, Price2018}, in contrast with the approach of REACH, rely almost entirely on the smoothness of the instrument response across frequency to avoid introducing spectral components in the foreground signals. Aiming at this instrument smoothness, EDGES restricted their frequency band to a bandwidth ratio of approximately 2:1 and operates scaled systems with a small overlapping band in order to cover the full frequency range. The joint statistical analysis approach used in REACH does not require that level of smoothness, and in some cases (see \textit{Methods / Foreground models and chromaticity correction}), certain frequency structure in the system response may be preferred, and therefore larger instantaneous frequency bands can be accessed (up to $\sim$ 3.5:1 with the conical log-spiral antenna).
    \item \textbf{Receiver calibration based on in-field measurements.} Extensive work has been carried out by experiment like EDGES \citep{2012RaSc...47.0K06R} to optimize the calibration of the receiver electronics by using a switched calibrator system. This calibration, to date, has relied on very meticulous and precise measurements of the different electronic components of the receiver in a well controlled laboratory environment. However, once deployed, the time stability of the system and its sensitivity to environmental effects (e.g. temperature) can translate into an inaccurate calibration. To overcome this, REACH features an in-field measurement system, using a compact Vector Network Analyzer (VNA) integrated with the receiver to provide real-time monitoring of the antenna and receiver components. 
\end{itemize}

\section{System design}

In this section we briefly describe the hardware system for REACH. Further information about the system and its location can be found in the \textit{Methods} section.

\begin{figure*}
\centering
\includegraphics[width=17cm]{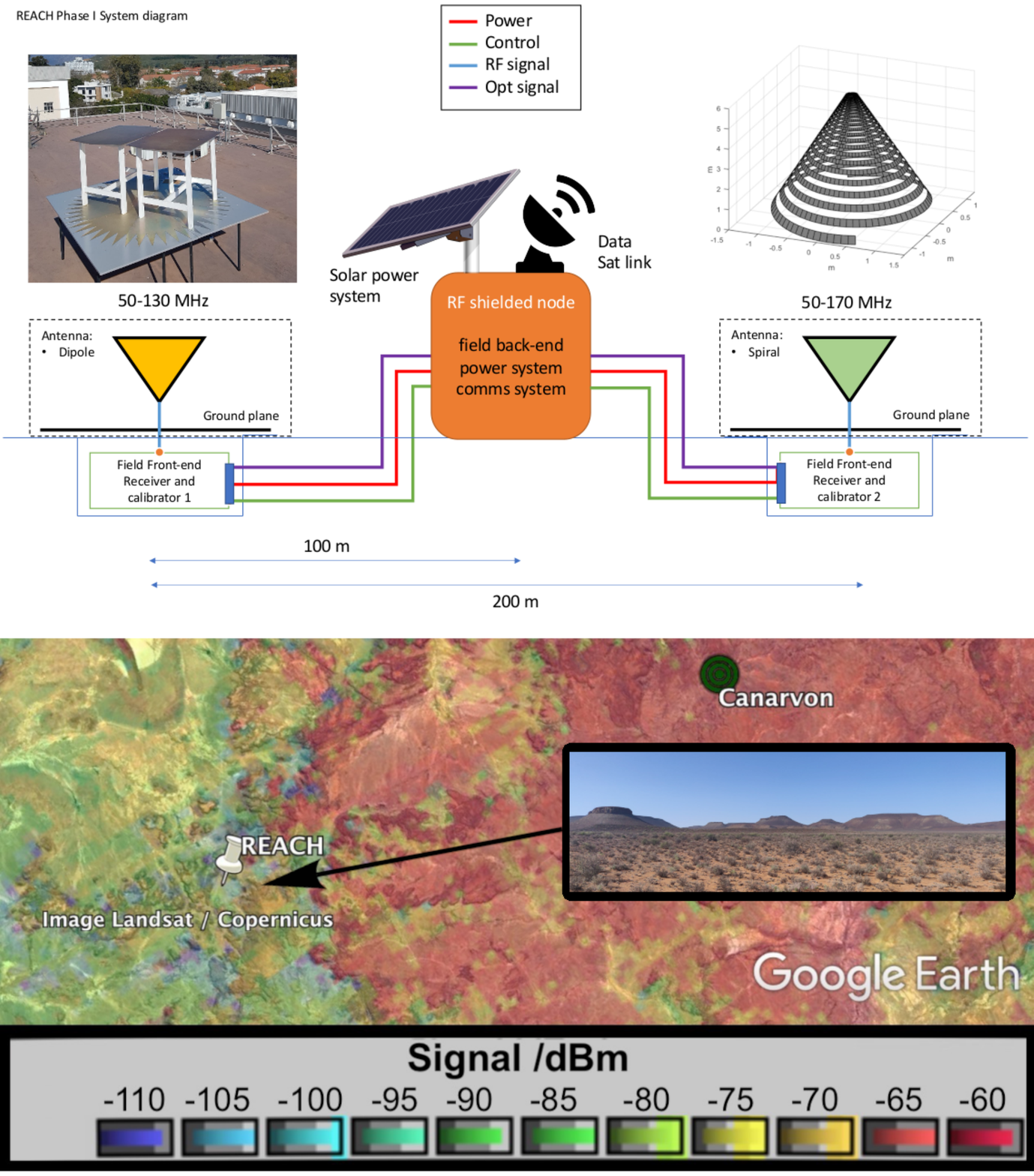}
\caption{Top: REACH Phase I field system featuring 2 independent radiometers. Bottom: Aerial satellite image  of the REACH site. Credit: www.peralex.com. A photo of the REACH site in the Great Karoo semi-desert in South Africa is shown in the small window.}
\label{fig:system}
\end{figure*}

\begin{figure*}
\centering
\includegraphics[width=17cm]{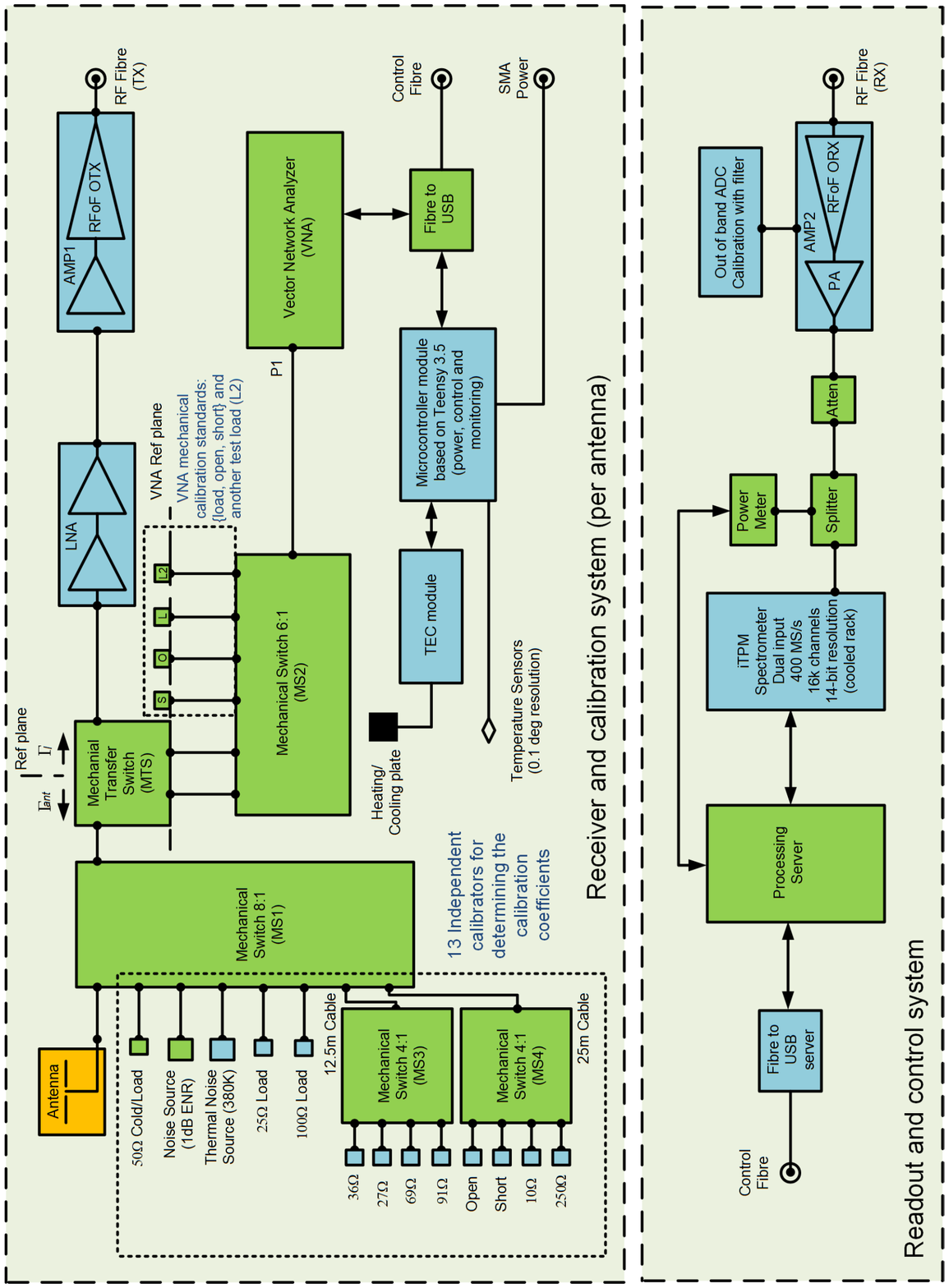}
\caption{Field and back-end hardware diagram (per antenna). The exact value of the calibration sources (left of the figure, connected to the mechanical switches) varies depending on the antenna connected to the receiver.  Green blocks represent off-the-shelf components, whilst blue are custom designs.}
\label{fig:RF_system}
\end{figure*}

In Fig. \ref{fig:system} we show a high level overview of the field system, while Fig. \ref{fig:RF_system} shows the details of the analogue and digital receivers, including the calibrator system. Subsection \textit{Methods / Antennas} describes the antennas that will be used in Phase I. 

REACH Phase I field system features 2 independent radiometers with different antenna designs (a hexagonal dipole antenna (50-130 MHz) and a conical log-spiral antenna (50-170 MHz)) as shown in Fig. \ref{fig:system}. REACH Phase I is a stand alone instrument powered from a solar power system and connected to the World via an internet satellite link. Its location in the Karoo semi-desert area of South Africa has been chosen based on several parameters including accessibility and low RFI. A radio propagation study, using the Longley-Rice propagation model, for total power in the FM band (88-108 MHz) radiated from the transmitter in Carnarvon across the topography of the Karoo Radio Astronomy Reserve is shown in Fig. \ref{fig:system}. The RF receiver system, incorporting the aforementioned in-field calibrator is followed by a high resolution spectrometer featuring 6 kHz-wide channels (for Radio Frequency Interference -RFI- excision) and is based on the SKA1-LOW Tile Processing Module (TPM) Field-programmable Gate Array (FPGA) board.

Subsection \textit{Methods / Receiver and Calibrator} explains the different hardware components in the analogue chain, including those used for its calibration. Subsection \textit{Methods / Digital back-end} is a high level overview of the digital high resolution spectrometer used in REACH. In subsection \textit{Methods / Site and RFI} we discuss the choice of location and the RFI environment. Finally, in subsection \textit{Methods / System Metrics} we summarize the main technical capabilities of REACH.

\section{Data analysis and Science predictions}

\subsection{Data analysis pipeline}

\begin{figure*}
\centering
\includegraphics[trim={0 2.5cm 0 0},clip,width=17cm]{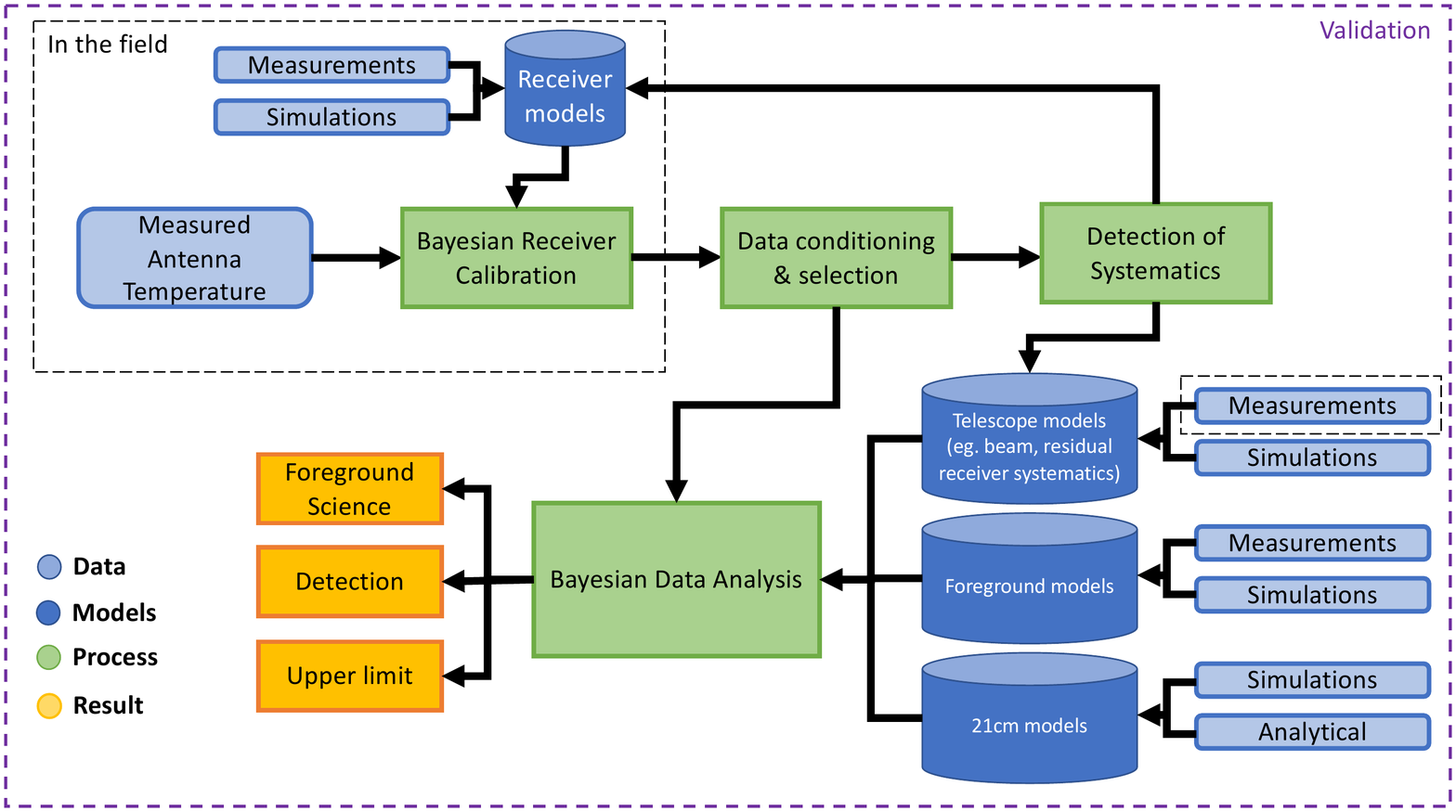}
\caption{Data analysis and calibration diagram. The diagram depicts the data flow around the joint Bayesian model fitting at the heart of REACH. A pre-calibration step takes place in the field, where the receiver electronics are calibrated using a Baesyian calibrator and a set of in-field measurements. Following this step, the data is transferred off-site, where after a pre-selection of high quality data, we perform an analysis in search for unknown (e.g. lacking a specific model) systematic signals remaining in the data. Finally, a joint model fitting process using Bayesian inference is performed on the data using pre-designed telescope, foreground and cosmological signal models. These models, specifically those associated with the telescope, will also be informed by direct measurements of, for example, the radio antenna.}
\label{fig:data_flow}
\end{figure*}

The data analysis pipeline for REACH Phase I is depicted in Fig. ~\ref{fig:data_flow}. After the receiver has been calibrated using in-field measurements (see Extended Data Fig. 2 and \textit{Methods / Bayesian receiver calibration} for more details), a pre-processing step of high quality data selection will be performed off-site (e.g. data from nights with a quiet ionosphere). This data will then be analyzed with a suite of tools to detect unknown systematic signals (see \textit{Methods / Detection of systematic errors} for more details). The result of this analysis will help inform the pre-designed telescope (see \textit{Methods / Instrument models} for more details), foreground (see Extended Data Fig. 3 and \textit{Methods / Foreground models and chromaticity correction} for more details) and cosmological models (see \textit{Methods / Cosmological models} for more details). For example, a potential outcome of the analysis of unknown systematic errors would be a new model component to capture this systematic signal. Then, the pre-calibrated and selected data will be used to jointly fit these models (see xtended Data Fig. 4, \textit{Methods / Bayesian data analysis and calibration} and \textit{Methods / Time- and antenna-dependant modelling} for more details). The output of this Bayesian fitting process will result in constraints on fundamental cosmological parameters as well as foreground parameters. This will allow us to interpret the models and produce the desired scientific results (see Subsection \textit{Science prospects} for more details).    

Going forward we expect to add further functionality to both the system (e.g. scaled antennas) and the analysis and calibration pipelines (e.g. statistical analysis of the receiver residuals during the joint data analysis step). We furthermore envisage the validation of any result using a comprehensive end-to-end simulation.

\subsection{Science prospects}

The primary science goal of REACH is noise-limited detection and observation of the evolution with redshift of the sky-averaged 21-cm hyperfine line emission from the neutral hydrogen that pervaded the intergalactic medium (IGM) during the Cosmic Dawn (CD) and the Epoch of Reionzation. The extreme challenge posed by strong foreground emission necessitates exquisite instrument modelling and data calibration in global 21-cm experiments. Ultimately, unless one can place strong priors on the parameters of models for plausible systematic effects in the data, the robustness with which constraints on astrophysics can be derived from the data is limited. A list of additional science outputs for the experiment is included in section \textit{Methods / Additional science outputs}.

\subsubsection{EDGES verification}

A $500~\mathrm{mK}$ absorption trough at $z=17$, consistent with the signal inferred by EDGES, is detectable with $\lesssim1/6$th of the observing time necessary for bright standard 21-cm signal models at the same redshift. As such, verifying the EDGES detection is the initial science goal of REACH. The EDGES analysis of their data utilises a flattened Gaussian parametrisation of the global 21-cm signal,
\begin{equation}
\label{Eq:FlattenedGaussian}
T_\mathrm{b}(\nu) = -A_\mathrm{flatG}\left(\frac{1-e^{-\tau e^{B_\mathrm{flatG}}}}{1-e^{-\tau}}\right) \ ,
\end{equation}
where,
\begin{equation}
B_\mathrm{flatG} = \frac{4(\nu-\nu_0)^2}{w^2}\log\left[-\frac{1}{\tau}\log\left(\frac{1+e^{-\tau}}{2}\right)\right] \ ,
\label{Eq:FlattenedGaussianB}
\end{equation}
and $A_\mathrm{flatG}$, $\nu_0$, $w$ and $\tau$ describe the amplitude and central frequency, width and flattening of the absorption trough, respectively. 

The presence in REACH data of a 21-cm signal that is consistent with that reported by EDGES can directly be constrained within the REACH analysis pipeline by performing a joint fit for the most probable model of the foregrounds and the global 21-cm signal model of the form shown in (\ref{Eq:FlattenedGaussian}), with priors on $A_\mathrm{flatG}$, $\nu_0$, $w$ and $\tau$ given by their posteriors in the EDGES analysis. As little as $\sim$ 3 hrs of integrated data out of approximately 30 hrs of observations, obtained over several nights (quietest ionosphere and lowest RFI) will enable us to verify the presence or absence of an EDGES signal of this form.

Beyond placing constraints on the presence of the global 21-cm signal inferred by EDGES directly, we will constrain the presence of a high amplitude global 21-cm signal, more generally, by fitting physically motivated exotic models for the 21-cm signal. If a 21-cm CD absorption trough with an amplitude in excess of $\sim$ 200 mK at $z=17$ exists in nature, it implies an increased differential brightness between the hydrogen spin temperature and the radio background, relative to standard cosmological models. As described in the \textit{Introduction}, this requires that either the radio background temperature is larger than expected, which would occur if there were an unaccounted for radio background in excess of the CMB \citep[][]{bowman18, feng18, ewall18, fialkov19, mirocha19, Brandenberger:2019, ewall20, reis20}, or, the hydrogen kinetic temperature, and, correspondingly, the spin temperature (following WF-coupling to the kinetic temperature), is cooled to below the level expected from adiabatic cooling alone \cite[][]{barkana18, Berlin, barkana18a, munoz18, Liu19}.

If REACH measurements find a 21-cm signal consistent with that inferred by the EDGES team, or, more generally, one in excess of standard astrophysical models for CD, we will fit 21-cm global signal simulations modelling each of these scenarios. In this case, we will place constraints on the presence of both standard astrophysical  and exotic sources of excess radio background emission  \citep{reis20}, as well as dark matter cooling of the hydrogen gas \citep{fialkov19}.

\subsubsection{Constraining a standard amplitude 21-cm signal}

If REACH measurements do not support the EDGES detection, as we continue to integrate down, the absence of a high amplitude signal will enable us to place stringent constraints on the amplitude of an excess radio background and on the physical mechanisms that can cool the hydrogen kinetic temperature to below the temperature expected from adiabatic expansion alone.

In this scenario, if continued integration leads to the detection of a standard astrophysical 21-cm signal, we will use 21-cm signal forward modelling consistent with standard astrophysical and cosmological assumptions to constrain astrophysical and cosmological parameters given our data.

\begin{figure*}
\centering
\includegraphics[width=17cm]{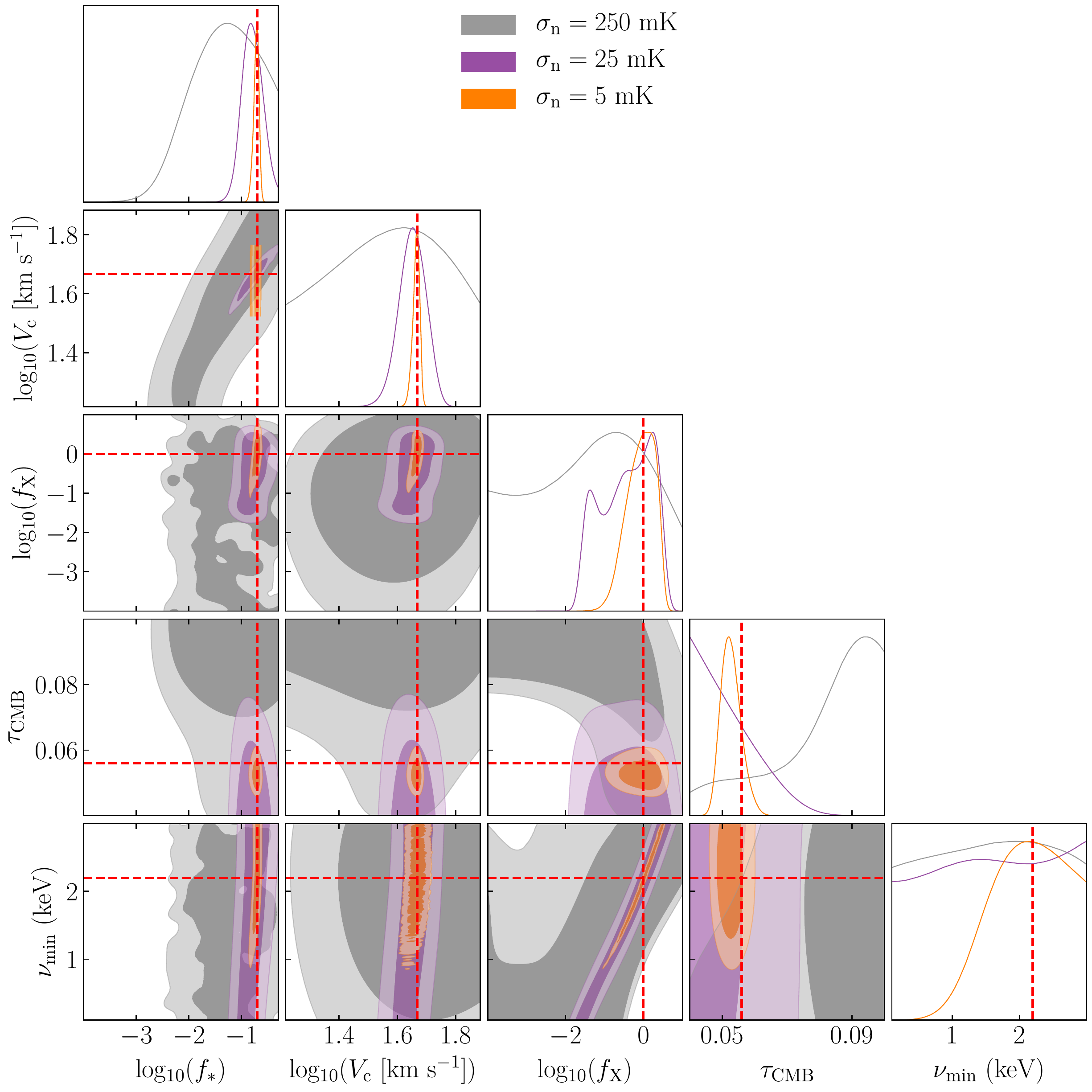}
\caption{Posterior probability distribution forecasts of constraints on 5 astrophysical parameters characterising the evolution of the 21-cm brightness temperature during Cosmic Dawn and the Epoch of Reionization. Forecasts are based on recovery of the astrophysical parameters of a fiducial 21-cm signal model (with input parameters shown with red dashed lines) from REACH data sets at three noise levels:  $250~\mathrm{mK}$ (grey), $25~\mathrm{mK}$ (purple), $5~\mathrm{mK}$ (orange). The parameters being constrained are $f_*$, the star formation efficiency; $V_\mathrm{c}$, the minimum virial circular velocity of star forming galaxies; $f_\mathrm{X}$, the X-ray efficiency of sources; $\tau_\mathrm{CMB}$, the CMB Thomson scattering optical depth; and $\nu_\mathrm{min}$, the low-energy cutoff frequency of the X-ray spectral energy distribution ($\alpha$, the power law index of the X-ray spectral energy distribution and  $R_\mathrm{mfp}$, the mean-free path of ionizing photons in the IGM are fixed in this analysis to 1.3 and 30.0 Mpc respectively since they tend to have a smaller influence on the global 21-cm signal). 
}
\label{Fig:science_forecasts2}
\end{figure*}

Figure \ref{Fig:science_forecasts2} illustrates the level of constraints on the properties of the IGM and the first luminous sources that would be inferred from a detection of a fiducial global 21 cm signal in the REACH spectral band in the limit that instrumental noise is the only source of uncertainty on the signal in the data. In practice, correlation between the foreground and signal models will increase the uncertainties on recovered 21-cm signal parameters; thus, Fig. \ref{Fig:science_forecasts2} represents a best-case scenario for the precision to which astrophysical parameters can be constrained at the quoted noise levels. The constraints, so derived, are illustrated for three noise levels (for simplicity Gaussian white noise has been added to the signal in the simulated data set. However, a more realistic radiometric noise model is under development and will be employed for analysis of REACH data): $250~\mathrm{mK}$ (grey), $25~\mathrm{mK}$ (purple), $5~\mathrm{mK}$ (orange). These correspond to the expected noise levels in data sets comprising observations centered on low-foreground regions of the sky, away from the Galactic plane, in a $100~\mathrm{kHz}$ channel, at a reference frequency of $50~\mathrm{MHz}$, with total observation times of $\sim 1$, $100$ and $2500~\mathrm{h}$, respectively. We treat these values as illustrative of pessimistic, fiducial and optimistic scenarios, respectively, for the non-systematics-limited noise levels that will be obtained with REACH. We define our fiducial scenario here with reference to the $25~\mathrm{mK}$ RMS of data-minus-sky-model residuals in EDGES low-band data found in \citep{bowman18}.
 
In practice, the specific constraining power of a REACH data set will depend on the shape and depth of the 21-cm signal relative to the noise on the data. Nevertheless, Fig. \ref{Fig:science_forecasts2} is illustrative of the general properties of the IGM and the first luminous sources that can be well constrained by data in the REACH band, with precision constraints on $f_*$, $V_\mathrm{c}$, $f_\mathrm{X}$, $\tau_\mathrm{CMB}$ and $\nu_\mathrm{min}$ obtained in the high signal-to-noise regime.

The 21-cm signal model in Fig. \ref{Fig:science_forecasts2} uses the \texttt{globalemu} global signal emulator \citep{bevins2021globalemu} of simulations by \cite{visbal12, fialkov14, cohen:2017, fialkov19,reis20}; however, alternate global signal modelling tools such as ARES \citep{Mirocha2014} are also available for use to the same end (see \textit{Methods / Cosmological models} for more details).

\section{Conclusion}
The REACH experiment has been conceived to provide a confident detection and subsequent analysis of the sky-averaged 21-cm signal from primordial hydrogen, a signal emitted at the time of the birth of the first stars and the epoch of reionization. Featuring a suite of Bayesian analysis and calibration techniques, the experiment improves over current efforts and their sensitivity to hardware systematics, by exploiting a joint analysis of the cosmological signal, the contaminating foregrounds and the instrument itself. This joint analysis provides us with several advantages, including access to the Bayesian evidence for the data and models as well as better understanding of the correlations between the different signal components. Two wideband radio antennas will observe simultaneously the southern hemisphere sky from the base at the Karoo radio reserve in South Africa. As shown in the manuscript, the use of simultaneous observations with two different antennae helps us enhance the precision of the detection as well as remain more insensitive to specific hardware related systematics. With a wide redshift coverage of 7.5-28, REACH will also use a novel in-field calibration technique using Bayesian parameter estimation of calibration coefficients. This approach allows us to further isolate systematic and hardware related features from spectral features belonging to the sky signal. In this paper we reported the experiment aims, system design, data pipelines and science predictions, highlighting the ability of the experimental approach to produce confident constrains of cosmological models.

\small

\appendix

\section{Methods} \label{Methods}

\subsection{Exotic interpretations of the EDGES signal}

In order to explain the depth of the feature reported by the EDGES collaboration, several exotic theoretical interpretations have been argued by the community. A solution can be obtained by achieving a stronger contrast between the spin temperature of the 21-cm transition and the temperature  of the radio background radiation at the intrinsic wavelength of 21 cm. Several physically motivated scenarios have been described in the literature, of which the most popular include: over-cooling of the hydrogen gas beyond the adiabatic cooling limit via interactions with cold dark matter  \citep{barkana18, Berlin, barkana18a, munoz18, Liu19} or producing extra high redshift  radio background contribution  in addition to the omnipresent CMB \citep{fixsen11, bowman18, feng18, ewall18, dowell18, fialkov19, mirocha19, ewall20, reis20}. 
In the former case, the over-cooling is mediated by millicharged cold dark matter (mDM) particles  interacting with ordinary matter and draining excess energy from the gas. Taking into account existing astrophysical, cosmological and particle physics constraints, recent studies indicate that the mass of a mDM particle should be between 10 MeV and a few hundreds of GeVs, with a dark matter energy density fraction between $10^{-8}$ and 0.004  in order to explain EDGES \citep{Liu19}. In the alternative scenario of an excess radio background,  anomalously  bright high-redshift astrophysical sources,  a thousand times stronger than the familiar radio loud active galactic nuclei  (AGN) \citep{urry95, biermann14, bolgar18, ewall18, ewall20} or star-forming galaxies \citep{condon92, jana19, reis20}, are required in order to sufficiently raise the radio background and explain the EDGES detection. Alternatively extra radio background could  be produced by exotic processes such as radiative decay of particles, annihilating dark matter or super-conducting cosmic strings \citep{Bolliet:2020, Brahma:2020, Fraser:2018, Pospelov:2018, Brandenberger:2019, caputo2020edges}. Other potential explanations include dark matter spin-flip interactions with electrons through a light axial-vector mediator directly inducing a 21-cm signal \citep{dhuria2021strong}.

\subsection{EDGES data re-analysis}

The spatial-spectral structure of the sky radiation, when convolved with the spatial-spectral antenna beam, produces spectral structure not easily modelled by a low order polynomial. A simulated analysis of this effect is provided in the \textit{supplementary information document - Figure 5}.

 It has been hypothesized \citep{2018Natur.564E..32H, 2019ApJ...880...26S, 2020MNRAS.492...22S,Bevins2020} that these kind of hardware residual systematic signals could be, at least in part, the explanation for the anomalous absorption profile published by the EDGES team \citep{bowman18}. The EDGES team described in detail in that publication their process to calibrate the foreground signal and hardware systematics, and they presented a comprehensive set of validation observations to support their choice of models and calibration. In that work a physically motivated polynomial-based model is used to calibrate the foreground signal after the beam chromaticity has been corrected for. However, while physically motivated, this foreground modelling used non-physical parameters and could be susceptible to model hardware systematics and potentially the cosmological signal itself, as shown in \citep{2018Natur.564E..32H}.
 
 In Extended Data Fig. 1 top-left we demonstrate that if the foreground models are restricted to physical parameters \citep{2018Natur.564E..32H} (purple) versus unrestricted foregrounds (orange) the detected signal from the EDGES public data could look quite different. Furthermore, Extended Data Fig. 1 bottom-left shows the best fit for the EDGES public data by using standard astrophysical models of the cosmological signal \citep{cohen19} instead of a flat Gaussian model as is used in \citep{bowman18}. The right column of Extended Data Fig. 1 shows the residuals after subtracting the posterior average foreground and signal models from the EDGES public data in each case. The root mean square values of the residuals are 0.023 K for the case of unrestricted foregrounds and a flattened Gaussian signal, 0.122 K for the case of restricted foregrounds and a flattened Gaussian signal, 0.077 K for the case of unrestricted foreground and a \texttt{21cmGEM} signal and 0.215 K for the case of restricted foregrounds and a \texttt{21cmGEM} signal. It is worth noting however that in \citep{bowman18} a series of tests are presented showing the insensitivity of the result to field hardware modifications.

\subsection{Antennas}

In REACH Phase I observations will be performed with two different antennas, in overlapping frequency bands, to detect and isolate antenna hardware systematics. These antennas will be installed on top of a 20 x 20 m metallic ground plane with serrated edges in order to minimize edge reflections. A multi-level design approach is followed in the development of the antennas. At the top level the general topology is down-selected, and more detailed designs are performed at lower levels of the most promising set of antenna types.  

The first level of down-selection is informed by a set of figures-of-merit (FoMs) describing the chromaticity of the antenna beam, as well as the impedance frequency response. A smooth beam is enforced by minimising the variance of the chromaticity factor $C(t,\nu)$. The original goal of this factor \citep{Mozdzen2017, Mozdzen2018} is to correct for the spectral structure introduced by the beam with respect to a given reference frequency within the band. It is defined as
\begin{equation}\label{eq:FoMchrom}
C\left(t,\nu\right) = \frac{\int_{\Omega}T_\mathrm{sky}\left(t,\nu_\mathrm{ref},\Omega\right)D\left(\nu,\Omega\right)\mathrm{d}\Omega}{\int_{\Omega}T_\mathrm{sky}\left(t,\nu_\mathrm{ref},\Omega\right)D\left(\nu_\mathrm{ref},\Omega\right)\mathrm{d}\Omega},
\end{equation}
where 
\begin{equation}
T_\mathrm{sky}\left(t,\nu_\mathrm{ref},\Omega\right) = \left[T_\mathrm{base}\left(t,\Omega\right) - T_\mathrm{CMB}\right]\left(\frac{\nu_\mathrm{ref}}{\nu_\mathrm{base}}\right)^{-\beta} + T_\mathrm{CMB}
\end{equation}
and $D\left(\nu,\Omega\right)$ is the antenna directivity, $\nu_\mathrm{ref}$ is the reference frequency, $\Omega$ is the direction on the sky, $T_\mathrm{base}\left(t,\Omega\right)$ is an all-sky base map at frequency $\nu_\mathrm{base}$, $t$ is the time of observation, and $T_\mathrm{CMB}$ is the CMB temperature. For the impedance response, we sought to maximise the frequency bandwidth where reflections into a 50~Ohm load is below $-10$~dB in magnitude, while enforcing the lowest reflection levels occur in the $50-170$~MHz band. Additionally, we prefer antennas with smoother impedance variation over frequency, as these are generally easier to model in computational electromagnetics (CEM) solvers.

The point of simple modelling is also considered in the down-selection process. Smaller antennas, with mechanically simpler structures, are generally easier to manufacture and model accurately. We expect better correlation between simulations and measured results for simple antennas than more complex structures, and thus more accurate feature extraction due to small geometric and material variations.

Once a few promising candidate antennas have been identified by the process above, a more complete pipeline analysis (briefly described in \textit{Methods / Data analysis driven antenna selection}) is performed on the nominal structures to estimate the likelihood and goodness of a simulated signal detection process. Since these analyses are extremely time consuming, they are not used in the direct optimization loop. Results from the pipeline simulations identified the conical log-spiral and horizontal hexagonal dipole antennas (see Fig. ~\ref{fig:system}) as the most promising structures. While the spiral antenna has a broader operating bandwidth than the dipole, it is mechanically more complex. The substantial difference in radiating mechanisms between these antennas, however, make them an attractive pair for REACH where we will use both antennas to identify and isolate hardware systematics in the analysis pipeline. We also note that the hexagonal dipole antenna shown in this manuscript is similar in type to the square dipole antenna used by the EDGES experiment.

The final antenna geometry is fine-tuned in a detailed optimization loop, where a parametric model of the main response features (both beam and impedance) is simultaneously extracted as a function of small variations in geometry around the nominal values. Details of this process are reported in \citep{cumner2021radio}.

\subsection{Data analysis driven antenna selection}

The different antennas under consideration were analyzed using a simulated version of the Bayesian data analysis pipeline, which is described in Section \ref{sec:data_pipeline}, as one of a number of Figures of Merit used to decide upon antenna designs (see Fig. ~\ref{fig:data_flow}). This assessment was based on our ability to reconstruct a range of injected mock 21-cm signal in simulated data with both a high degree of statistical confidence and a small RMS error with respect to the injected signal for different simulated antenna patterns.

The REACH data analysis pipeline is designed to correct for systematic distortions of data due to chromaticity of the antenna. However, certain structures of chromatic distortion can be more difficult to correct for than others. Therefore, testing how well a range of signals are recovered from simulated data, which includes modelling of the distortions from the antenna, using the pipeline as part of the design process helps identify what antenna designs are more suited to the experiment. This process is described in detail in \cite{anstey2021informing}. A summary of this selection strategy is provided in the \textit{supplementary information document - Figure 6.}

\subsection{Receiver and calibrator}

The REACH radiometer uses the noise waves formalism \cite{meys} to determine critical calibration coefficients.  Its primary aim is to achieve milli-Kelvin level calibration of the sky signal by correcting for the antenna mismatch and receiver response including the analogue-to-digital converter.  This is achieved by careful design of the RF components in the chain and the use of very high quality calibration sources at known temperatures. 

The REACH radiometer presents the ultimate in instrumental calibration capability because it relies purely on field gathered data which is fully autonomous requiring no physical interaction once installed, meaning that calibration and observation can happen concurrently.  The details of the  noise waves along with the Bayesian receiver calibration pipeline we have developed which employs conjugate priors to compute coefficients quickly in the field, can be seen in \textit{Methods / Bayesian receiver calibration}. A full description can be found in \citep{reachBayes}. Furthermore, the radiometer shown in Fig.~\ref{fig:system} can easily incorporate many "calibration sources" and is currently envisaged to have 13 independent calibrators all feeding data into the receiver calibration pipeline.

The calibration system employs mechanical switches which have very low loss (typically 0.01~dB in this band) and better than 100~dB isolation.  A transfer switch allows a VNA to measure the reflection coefficients of the sources and the low-noise amplifier (LNA).  An on-board micro-controller facilitates this process along with controlling the environmental temperature.  All signals (control and RF) are transported via RF-over-fibre cables to avoid interference and extra loss to a readout system which converts signals back into RF and digitises them.  The calibration sources used offer strategic sampling of the noise waves as a function of impedance and go much further than a standard open or shorted cable. More information on this is provided in the \textit{supplementary information document - Figure 1}.

In a normal observation REACH will rely on Dicke switching to observe the sky, an ambient load and a noise source on regular intervals of 10-30 seconds.  Using these three power spectral densities measured with the spectrometer we can calibrate out the effects of the receiver system.

The first major component in the RF chain is the LNA which is in the field unit under the antenna.  It is designed to have a very flat spectral response both in terms of S-parameters and noise. This component is followed by AMP1 which converts the signal into an optical signal. The optical signal from the field unit is converted back into RF with AMP2 in the back-end, where out of band noise (DC-25MHz) is also injected for conditioning of the spectrometer.  Finally the signal is split and fed to both a high resolution power meter and the spectrometer running side by side.  This offers additional information which can be used in the calibration process.  The passband response of the REACH receiver system along with the measured LNA response is provided in the \textit{supplementary information document - Figure 2}.

\subsection{Bayesian receiver calibration}
\label{rxcal}

REACH uses a novel calibration algorithm developed from the formalism introduced in the EDGES experiment \citep{bowman18}. The calibration strategy follows the method established by Dicke \citep{dicke} to characterise systematic features of radio-frequency instruments through measurements of multiple calibration standards in order to determine `noise wave parameters'; $T_{\mathrm{unc}}$, $T_{\mathrm{cos}}$, $T_{\mathrm{sin}}$, $T_{\mathrm{NS}}$ and $T_{\mathrm{L}}$ \citep{dicke, meys}. These noise wave parameters describe reflections arising at connections within the experimental apparatus which re-enter the receiver along with the measurement. The use of noise wave parameters allows for the derivation of a relationship between the calibrated temperature of any source, $T_\mathrm{source}$, and power spectral densities measured by the receiver \citep{reachBayes}

\begin{equation}
  \label{eqn:caleqn}
  \begin{aligned}
  T_\mathrm{NS} \left( \frac{P_\mathrm{source} - P_\mathrm{L}}{P_\mathrm{NS} - P_\mathrm{L}} \right) + T_\mathrm{L}&= T_\mathrm{source}\left[ \frac{1-|\Gamma_\mathrm{source}|^2}{|1-\Gamma_\mathrm{source}\Gamma_\mathrm{rec}|^2} \right] \\
  & + T_\mathrm{unc}\left[ \frac{|\Gamma_\mathrm{source}|^2}{|1-\Gamma_\mathrm{source}\Gamma_\mathrm{rec}|^2} \right] \\
  & + T_\mathrm{cos}\left[ \frac{\operatorname{Re}\left(\frac{\Gamma_\mathrm{source}}{1-\Gamma_\mathrm{source}\Gamma_\mathrm{rec}}\right)}{\sqrt{1-|\Gamma_\mathrm{rec}|^2}} \right] \\
  & + T_\mathrm{sin}\left[ \frac{\operatorname{Im}\left(\frac{\Gamma_\mathrm{source}}{1-\Gamma_\mathrm{source}\Gamma_\mathrm{rec}}\right)}{\sqrt{1-|\Gamma_\mathrm{rec}|^2}} \right], \\ 
  \end{aligned}
\end{equation}
where $P_{\mathrm{source}}$, $P_{\mathrm{L}}$ and $P_{\mathrm{NS}}$ are measured power spectral densities of the receiver input, internal reference load and internal reference noise source respectively. $\Gamma$ represents measured reflection coefficients of the same source and references. We may condense our calibration equation into a linear relationship for simplicity
\begin{equation}
  T_\mathrm{source} = X_\mathrm{unc}T_\mathrm{unc} + X_\mathrm{cos}T_\mathrm{cos} + X_\mathrm{sin}T_\mathrm{sin} + X_\mathrm{NS}T_\mathrm{NS} + X_\mathrm{L}T_\mathrm{L}. 
\end{equation}

Taking advantage of the linear form of our equation, we may group our terms into a matrix containing our data, $\mathbf{X}$, and a matrix of calibration parameters, $\boldsymbol{\Theta}$

\begin{align}
  \textbf{X} &\equiv \begin{pmatrix} 
  X_\mathrm{unc} \quad 
  X_\mathrm{cos} \quad
  X_\mathrm{sin} \quad
  X_\mathrm{NS} \quad
  X_\mathrm{L} \end{pmatrix}\nonumber\\
  \boldsymbol{\Theta} &\equiv \begin{pmatrix} 
  T_\mathrm{unc}\quad
  T_\mathrm{cos}\quad
  T_\mathrm{sin}\quad
  T_\mathrm{NS}\quad
  T_\mathrm{L}\end{pmatrix}^\top.
\end{align}

Our calibration equation can now be solved using a linear regression model

\begin{equation}
    T_\mathrm{source} = \mathbf{X}\boldsymbol{\Theta}+\sigma,
\end{equation}

with our noise term, $\sigma$. This allows for a joint solution of all terms instead of an iterative approach as used in previous experiments such as EDGES \citep{calpap}.

The application of conjugate priors within our Bayesian methodology enables our algorithm to be many orders of magnitude faster than techniques that use full numerical sampling via Markov chain Monte Carlo methods over many parameters. This allows for an in-place calibration with the data acquisition instead of relying on off-site measurements. Individual noise wave parameters are optimised using a gradient descent algorithm rather than applying a blanket fit multi-order polynomial to all noise wave parameters. Correlation between noise wave parameters is also considered in the derivation of their values as shown in Extended Data Fig. 2 \citep{reachBayes}. A schematic of the calibration algorithm is shown in Extended Data Fig. 2. Preliminary trials using this technique, applied to eight calibration standards, successfully derive the temperature of a $50 \ \Omega$ load resulting in an RMS error of 8 mK between calibrated and measured temperatures, well within a $1\sigma$ noise level and comparable to the calibration accuracy of the EDGES experiment \citep{reachBayes, calpap}.

\subsection{Digital back-end}

The digital back-end (or REACH spectrometer) is based on the FPGA digitalizer and beamformer board developed for the SKA1 Low Frequency Array \cite{10.1117/12.2232526}. This board, known as TPM, hosts sixteen high-performance Analog Devices AD9680 14 bit dual-channel Analogue-to-Digital Converters (ADCs) and two Xilinx Ultrascale FPGAs. The board has been successfully used in the context of the Aperture Array Verification System for SKA \cite{Naldi:tpm:2017} and in several other instruments \cite{melis:pharos:2020, Locatelli:frbs:2020, Magro:mexart:2019}. Together with its firmware and software libraries \cite{Magro:tpm_mc:2017}, the TPM provides a platform for fast development of radio-astronomy digital back-ends. In this context, the auxiliary functions, such as communication over Gigabit Ethernet for Monitoring and Control and data acquisition, are reused with minimal modifications, while the FPGA firmware is customized to support the specific application requirements. Specifically, a high-performance digital spectrometer was developed, where each FPGA processes a single digitised RF signal. Each TPM can thus process two RF-signals, with the unused ADCs being powered down to save power. This arrangement allows for the implementation of a full floating-point Digital Signal Processing (DSP) pipeline which includes a polyphase filterbank (PFB) and power integrator. The key parameters of the TPM spectrometer are summarised in the \textit{supplementary information document - Table 1}. 

We note that an important aspect of the digital back-end is the channel isolation, which we have measured to be 90 dB (side-lobe rejection at the adjacent channel center) as shown in the \textit{supplementary information document - Figures 3 and 4}. This value is especially relevant in the context of RFI excision. While deeper RFI measurements are needed, our initial measurements \citep{Josaitis} and experience with co-located experiments (eg. HERA) indicate that this value could be good enough already.

The PFB uses a weighted overlap-add architecture supporting a total number of 229,376 tap coefficients, which can be download to the FPGA, allowing for the use of different weighting functions without re-compiling the FPGA firmware. Power spectra are accumulated over a programmable number of Fast Fourier Transform (FFT) frames which is set to a corresponding integration time of about $\sim$ 1 s. Accumulated spectra are then transmitted to the processing server where they can be accumulated further.  
The back-end control software is responsible for the correct, timely and safe operation of all the controllable hardware making up the on-site deployed system. It is responsible for configuring, initialising and controlling the required components for the correct running of observations defined by the operator. An observation includes a number of steps; receiver calibration, continuous source switching, spectra accumulation and hardware monitoring being some of the primary steps. Upon observation completion, the generated output files can then be transferred off-site through the satellite network link.

\subsection{Site and RFI}
\label{sec:site}

Conventional approaches to the analysis of 21-cm data fundamentally assume that spectrally-smooth astrophysical foregrounds can be discriminated from the predicted spectral structure of the 21-cm signal. Highly chromatic, terrestrial RFI is several orders of magnitude brighter than the spectral foregrounds, whose brightness itself already poses significant challenges to the 21-cm analysis pipeline. The spectral structure of RFI, if not carefully removed, could be considered as part of the spectral structure of the EoR signal itself. In an effort to mitigate the effects of RFI, the REACH instrument will be deployed at a radio quiet site in the Karoo Radio Astronomy Reserve (‘the Reserve’) near the town of Carnarvon in the Great Karoo semi-desert in South Africa. The Reserve is shared with several other radio telescopes including the Square Kilometre Array mid-frequency core \cite{Dewdney_SKA}, MeerKAT \cite{Jonas_MeerKAT}, HERA \cite{DeBoer_2017}, and HIRAX \cite{HIRAX_SPIE}, and offers critical support infrastructure to our experiment not always available at radio-quiet sites, such as well-maintained roads, on-site maintenance and engineering staff, and controlled access and entry. The site is located within 7 hours drive of Cape Town, making it feasible for the Collaboration to routinely commission the instrument and also offer access to academic and manufacturing resource hubs.

Since most of the other instruments at the Reserve (especially the flagship SKA-mid telescope) operate at higher frequencies, less severe restrictions are in place for radio transmissions at REACH operating frequencies in the surrounding region. As such, care was taken to select a site with minimal RFI at the REACH operating frequencies. The most problematic sources of RFI are FM radio transmitters serving surrounding communities, which operate in the band 88-108$~$MHz. After two extensive RFI measurement surveys, where a number of potential sites (informed by Longley-Rice propagation models with $1~$km$^2$ resolution of the FM signals over the terrain of the Reserve shown in Fig. ~\ref{fig:system}) were investigated, the final site was selected on the basis of a combination of low RFI, access to a nearby road, and favorable topography. Details of the measured RFI results for several sites, as well as a full description of the methodology and measurement calibration, are provided in \cite{Josaitis}. The site ($30^\circ50'16.75''~$S; $21^\circ22'27.22''~$E), depicted in Fig.~\ref{fig:system}, is a large flat plane of approximately 4 km in diameter, surrounded on all sides by mesas and hills. Although the interference from FM radio transmitters is reduced at this site, it is still present at a low level and several narrow bands must be masked out in the final data analysis pipeline. The interfering power levels are low enough to be of no concern for saturating the receiver.

\subsection{High-level system metrics}

The REACH Phase I instrument features 2 radiometer systems. The first one using a wideband hexagonal dipole antenna covering a frequency band of 50-130 MHz, with zenith directivity of 7.0 dBi and a radiation efficiency of 0.9. The receiver noise temperature equals 600K for this radiometer. The second radiometer system uses a conical log-spiral antenna, covering a frequency band of 50-170 MHz, with zenith directivity of 11.0 dBi and a radiation efficiency of 0.98. The receiver noise temperature equals 600K for this radiometer as well.

With these parameters, the minimum required integration time to detect an EDGES like signal (0.5 K) with a signal-to-noise ratio of 10:1 on 100 kHz coarse channels would be $\sim$ 6.5 hrs with the dipole system and 7.3 hrs with the conical log-spiral system. A standard 0.15 K signal centred at 100 MHz would require $\sim$ 29 hrs with the dipole system and 31 hrs with the conical log-spiral. Meanwhile, an exotic deep signal (0.5 K) centred at 130 MHz would require only $\sim$ 1.3 hrs with the dipole and a very similar number with the conical log-spiral. These simplistic estimates are purely based on a radiometer equation calculation and meant to give a first order estimate of the required integration time for the REACH system depicted in this manuscript. In reality, observations will need to last at least an order of magnitude longer, since for example we will need to flag and blank a significant amount of low quality data (e.g. contaminated by RFI) and to spend part of the observation time calibrating the receiver unit. 

\subsection{Instrument models}

A fundamental component of the data analysis pipeline is the use of parametrized physically motivated instrument models accounting for the response of the radio telescope to the sky signal across angular space, time and frequency. Especially important is the ability to account for environmental effects and interaction with the environment (e.g. the soil underneath the antenna). Within REACH, we are developing these models using a suite of tools, that include: full wave electromagnetic simulations, Neural Network emulators, lab measurements of components, measurements of scaled prototypes in control environments (e.g. radio anechoic chambers), in-field measurements using Unmanned Aerial Vehicle platforms \citep{UAV_SKA}, etc. 

The impact of soil and ground plane truncation on the electromagnetic properties of the antenna will be taken into account using
the method described in \citep{8917790}. Using this method, the S-parameters and radiation patterns
of the antenna do not suffer from a lack of accuracy resulting from the use of the method of images. In \citep{8917790}, a full wave solver is developed to solve the Maxwell's equations in their integral form.  
This method is based on a spectral formulation of the Method of Moments, which uses inhomogeneous plane waves to describe the interactions between the antenna and the finite ground plane. In a nutshell, the antenna equivalent currents are decomposed into a spectrum of inhomogeneous plane waves (IPWs) that is integrated to obtain the spatial electric field radiated by the antennas on the ground plane. This field is then integrated on the ground plane equivalent currents and the MoM linear system of equations is solved.
The field incident to the ground plane can be expressed as follows:
\begin{equation}
    \begin{split}
        \vec{E}_p(x,y,z)  & =  \frac{-j k \eta}{{(2 \pi)}^2} \int \! \! \! \int   F_p(k_x,k_y)   \\ 
 & \times \frac{e^{-j (k_x x + k_y y - k_z z) }}{2 j k_z} \left[ \hat{\textbf{e}}_p^d 
+ \Gamma_p(\beta) \ \hat{\textbf{e}}^u \right] \  dk_{x} \ dk_{y} ,
    \end{split}
\end{equation}
where $F_p(k_x,k_y)$ is the antenna radiation pattern, $(k_x,k_y,k_z)$ are the components of the wave vector, $k$ is the wavenumber, $\eta$ is the medium impedance,
$\hat{\mathbf{e}}_p$ is the polarization vector of polarization $p$ (either TE or TM) and $\Gamma_p(\beta)$ is the reflection coefficient due to the presence of the soil.
This is an efficient formulation to compute the interactions between an antenna and a ground plane with electrically small diameter ( $< \lambda_0$, with $\lambda_0$ the free space wavelength).
In [1], an accelerated formulation based on analytical Hankel transforms is derived to efficiently handle the interactions
with large ground planes. This second formulation is efficient to compute the intermediate electric field ( distances greater than $ \lambda_0$). Combined with the IPWs method, the electric field can be rapidly evaluated everywhere on the ground plane. Finally, these models will be informed with measurements of the soil layers using a Ground Penetration Radar system.

\subsection{Bayesian data analysis and calibration}

Theoretically, the temperature of the sky $T_\mathrm{sky} =T(\Omega,\nu,t)$ is a function of angle $\Omega$, frequency $\nu$ and time $t$. We work in earth coordinates, with the sky rotating over us providing some of the time variation, and the other portion arising from time-dependent foregrounds such as the ionosphere. We therefore decompose the raw sky signal into the global signal $T_\mathrm{g}(\nu)$ which is constant across the sky and time, the foregrounds $T_\mathrm{f}$ and statistically random components $N_\mathrm{sky}$
\begin{equation}
    T_\mathrm{sky}(\Omega,\nu,t) = T_\mathrm{g}(\nu) + T_\mathrm{f}(\Omega,\nu,t) + N_\mathrm{sky}(\Omega, \nu, t).
\end{equation}
This equation makes explicit the standard modelling split between elements of the system which we can model deterministically ($T_\mathrm{g}$ and $T_\mathrm{f}$) and elements which at best we can model probabilistically ($N_\mathrm{sky}$).
In practice a global experiment collects data, $T_\mathrm{obs}$, from the full sky convolved with an antenna directivity $D$:
\begin{equation}
    T_\mathrm{obs}(\nu,t) = \int T_\mathrm{sky}(\Omega,\nu,t) D(\Omega,\nu) \mathrm{d}\Omega + N_{T_\mathrm{obs}}(\nu,t).
\end{equation}
This process introduces its own random noise $N_{T_\mathrm{obs}}$ associated with the calibrator~\cite{reachBayes}. In this notation we take a broad definition of beam modelling $D$ which for example in addition to taking hardware into account can also include the effects of the ionosphere through an augmented angle argument $\Omega$~\citep{Shen2020}. Assembling these pieces we have:
\begin{gather}
    T_\mathrm{obs}(\nu,t) - T_\mathrm{g}(\nu) - \int T_\mathrm{f}(\Omega,\nu,t) D(\Omega,\nu) \mathrm{d}\Omega = N(\nu,t), \\
    N(\nu,t) \equiv N_{T_\mathrm{obs}}(\nu,t) + \int D(\Omega,\nu) N_\mathrm{sky}(\Omega,\nu,t) \mathrm{d}\Omega.
\end{gather}
where the random portions have been combined into a single random noise variable $N$. A common next step is to then ``integrate'' this over time, but this is only strictly necessary if data compression is required for storage purposes. In practice, most likelihoods will perform an effective integration step and to speed up computation one can usually do this sum early on, but it is important to conceptually separate convenience from necessity for now.

The statistical approach is to then use the random distribution of $N$ to generate a likelihood. For simplicity and concreteness if we assume that $N$ is Gaussian and uncorrelated in time and frequency with noise level $\sigma$, the probability of observing the antenna data, $T_\mathrm{obs}$, is
\begin{align}
    P(T_\mathrm{obs}|T_\mathrm{g},T_\mathrm{f},D) =& \prod_{\nu,t}\frac{1}{\sqrt{2\pi}\sigma}e^{-N^2/2\sigma^2}
    \quad\Rightarrow 
    \label{eqn:likelihood}\\
    \log P(T_\mathrm{obs}|T_\mathrm{g},T_\mathrm{f},D) =& -\sum_{\nu,t}\log{(\sqrt{2\pi}\sigma)} \nonumber\\&\qquad+ \frac{1}{2\sigma^2}\left(T_\mathrm{obs}-T_\mathrm{g} - \int T_\mathrm{f} D \mathrm{d}\Omega\right)^2 \nonumber.
\end{align}
This approach can of course be extended to a more sophisticated noise model (e.g. to a $\sigma$ model with varying frequency $\nu$ or correlations, or using an explicit calibrator derived setup). It is also worth noting that from a statistical perspective, this is what is being implicitly assumed whenever one performs a least-squares fit between model and data.
 For our REACH setup, after calibration the noise model is in fact a Student-$t$ distribution, rather than a Gaussian, and our final pipeline will use Bayesian model comparison for the final likelihood design and selection~\cite{ScheutwinkelA}.

The quantity in equation \ref{eqn:likelihood} is termed the likelihood $P(T_\mathrm{obs}|T_\mathrm{g},T_\mathrm{f},D)$, and is the cornerstone of both frequentist and Bayesian approaches. Both approaches then often begin by parameterising the unknown components: the global signal $T_\mathrm{g} = T_\mathrm{g}(\nu;\theta_\mathrm{g})$, the foregrounds $T_\mathrm{f} = T_\mathrm{f}(\nu,\Omega, t;\theta_\mathrm{f})$, and the directivity $D = D(\Omega,\nu;\theta_D)$, where $\theta_g$, $\theta_\mathrm{f}$ and $\theta_D$ are each vectors of parameters, and we denote all parameters as $\theta=(\theta_\mathrm{g}, \theta_\mathrm{f}, \theta_D)$. The likelihood can then be viewed as a function of these parameters:
\begin{multline}
    P(T_\mathrm{obs}|\theta) \equiv \\
     P(T_\mathrm{obs}(\nu,t)|T_\mathrm{g}(\nu; \theta_\mathrm{g}),T_\mathrm{f}(\nu, \Omega, t; \theta_\mathrm{f}),D(\Omega,\nu; \theta_D)).
\end{multline}

These parameterisations may be physical, phenomenological or non-parametric. For example, in the case of the global signal, we may wish to use signals generated by an emulator (e.g. \texttt{21cmGEM} \citep{cohen19} or \texttt{globalemu} \citep{bevins2021globalemu}) parameterised by physical parameters of the early universe, or a phenomenological flattened Gaussian, or a non-parametric free-form fit to the dip using a spline or polynomial approach~\cite{2019PhRvD.100j3511H}.

The Bayesian approach treats the parameters $\theta$ as unknown variables. After stating a prior $P(\theta)$, we proceed to compute the likelihood and evidence via Bayes theorem:
\begin{align}
    P(\mathrm{Data}|\theta) P(\theta) &= P(\theta|\mathrm{Data}) P(\mathrm{Data}).  \\
    \text{Likelihood} \times \text{Prior} &= \text{Posterior} \times \text{Evidence}.
\end{align}
Laid out in this non-conventional form~\citep{skilling2006}, Bayes theorem states that our inputs to inference are a likelihood $P(\mathrm{Data}|\theta)$ and prior $P(\theta)$, and our outputs are a posterior $P(\theta|\mathrm{Data})$ and evidence $P(\mathrm{Data})$. The posterior tells us our degree of knowledge of the parameters in light of the data, and can be used to marginalise out quantities which we are not interested in (nuisance parameters, such as the calibration or directivity details), and can also be used to produce forward inferences on quantities derived from these parameters, such as the global signal itself, or the foregrounds. The evidence allows us to perform Bayesian model comparison to determine the quantitative merits of a given set of parametric models. This is critical on two fronts. It first allows us to choose the best set of modelling assumptions, and, second, it allows us to determine the probability that there is a signal in the data in comparison to a fit where no global signal is included. Crucially evidences do this in a way that generates Occam's Razor as a theorem ~\citep{2021arXiv210211511H}.

Nested sampling ~\citep{skilling2006} is a robust tool to numerically sample the full posterior and calculate the Bayesian evidence. Nested sampling is well suited for navigating a-priori unknown complex posterior surfaces which may or may not have multiple posterior modes and non-trivial covariance structure between parameters. Such structures regularly occur when fitting sophisticated non-parametric models where there are potential hidden degeneracies between signal and foreground~\citep{2019PhRvD.100j3511H}. Our nested sampling implementation of choice is \texttt{PolyChord}~\citep{2015MNRAS.453.4384H,2015MNRAS.450L..61H} whose slice sampling based approach represents the state-of-the-art in nested sampling in high dimensional parameter spaces and has the unique capability of being able to exploit a fast-slow parameter hierarchy, which occurs naturally in the context of 21-cm modelling.

\subsection{\label{sec:data_pipeline}Foreground models and chromaticity correction}

Detailed data analysis processes are required in global 21-cm experiments such as REACH in order to separate the 21-cm signal from foreground emissions in the observing band. These foreground emissions come from many different sources, primarily galactic synchrotron and free-free emissions, as well as extragalactic point sources \citep{Shaver1999}. At the redshifts relevant to the global 21-cm signal, they can exceed the signal by up to four orders of magnitude. The first step in identifying the signal beneath these foregrounds is to exploit the fact that they are predominantly very spectrally smooth power law emissions, whereas the signal is not \citep{Pritchard2010,bowman2010, presley2015}. This would allow the signal to be identified by subtracting off smooth structure.

However, this process is made substantially more difficult by the presence of antenna chromaticity. The necessity of observing a wide frequency band makes the chromaticity in the pattern of the antenna very difficult to avoid. These changes in the antenna pattern with frequency then act to couple spatial variations in power on the sky into the frequency domain, resulting in non-smooth structure in the foregrounds that inhibits identification of the signal \citep{Bernardi2015, Bernardi2016, Rao2017,Singh2017, Nhan2017, monsalve17,Monsalve2018,Monsalve2019}. If these distortions are not accurately corrected for, it can result in residual systematics in the data that can prevent detection of the signal \citep{2018Natur.564E..32H, 2020MNRAS.492...22S, Bevins2020}.  

One way of performing this correction is divide the data by a correction factor $C\left(t,\nu\right)$, defined in equation \ref{eq:FoMchrom}, as described in \citep{Mozdzen2017, Mozdzen2018, monsalve17}. Implementations of this correction, however, make assumptions such that the sky at the frequency of the base map has the same spatial power distribution as at the reference frequency, that the spectral index of the foregrounds is uniform \citep{Mozdzen2017, Mozdzen2018} or that the simulated antenna pattern is an accurate model of the true pattern. These assumptions cannot be guaranteed in practice, which may result in this process leaving uncorrected-for systematics. 

More sophisticated techniques for the correction of chromaticity distortion have also been proposed, for example including a frequency-dependent sky brightness distribution \citep{monsalve17}.
Another proposed method exploits the possibility to model systematic effects using a single value decomposition analysis of simulated observations, possibly avoiding the need of accurate instrument models \citep{Tauscher2018,Rapetti2020,hibbard20}.

However, in keeping with the philosophy of the REACH project, we perform this correction, and the subsequent removal of foregrounds, by means of a new Bayesian data analysis pipeline that incorporates detailed physical modelling of the effect of chromaticity on the foregrounds \citep{anstey20}. This allows an understanding of exactly what systematics are being removed.

In practice, this process works by generating parameterised approximate models of the full sky at all observing frequencies. The simplest way to do this would be to scale a known all-sky map by a single uniform spectral index parameter to  the relevant frequencies. However, the spectral index actually varies across the sky \citep{Rogers2008,deOliveira2008, Guzman2011, Patra2015, Mozdzen2017, Mozdzen2018, Spinelli2020}, and if this is not accounted for, the resulting chromatic distortion models will not be accurate enough for the 21-cm signal to be identified. Therefore, we adopt a more detailed model that includes spectral index variation by subdividing the sky into $N$ regions of similar spectral index, such as shown in Extended Data Fig. 3 for $N=6$. A separate variable spectral index parameter can then be assigned to each region and a known all-sky map scaled by the resulting coarse-grained spectral index map to create the necessary parameterised sky model.

This sky model, $T_\mathrm{f}\left(\Omega, \nu, \theta_\beta\right)$, parameterised by spectral index parameters $\theta_\beta$, can then be convolved with the antenna beam model, $D\left(\Omega, \nu, \theta_\mathrm{A}\right)$ as discussed above in a simulated observation, to produce a data model.

\begin{equation}
T_\mathrm{model}\left(\nu\right) = \int_{\Omega}D\left(\Omega, \nu, \theta_\mathrm{A}\right) T_\mathrm{f}\left(\Omega, \nu, \theta_\beta\right)\mathrm{d}\Omega.
\end{equation}
 
The result of this a foreground model which is parameterised by a physical property, the spectral index of the foreground emissions, and includes the chromatic distortions from the antenna as a part of the model. This allows the distortions to be fit from the data as part of the foreground rather than needing to be simulated and corrected for in advance.

We then propose to fit this foreground model to the data alongside a parameterised 21-cm signal model using the Bayesian nested sampling algorithm \texttt{PolyChord} \citep{2015MNRAS.453.4384H,2015MNRAS.450L..61H}. The foreground residuals and recovered 21-cm signals from fitting simulated data of a log-spiral and hexagonal dipole antennas are shown in the rightmost column of Extended Data Fig. 3. Extended Data Fig. 3 also shows the results of fitting the simulated data with a log-polynomial model, with and without chromatic corrections of the chromaticity factor $C\left(\nu\right)$ (equation (\ref{eq:FoMchrom})), for comparison.

Using a nested sampling algorithm gives access to the Bayesian evidence. This can be used to allow the data to inform how many parameters are needed in the foreground and instrument models and thus prevent the data being fit with a foreground model of too many parameters, which might otherwise obscure the 21-cm signal.
It also enables comparisons between models that include or do not include signal models, to quantify confidence in the presence of a signal, and comparisons between difference signal models.

By modelling the foregrounds and chromatic distortions together in a physically motivated manner, it enables the systematics that arise due to chromaticity to be well understood and accounted for. It also allows other sources of systematics, such as polarisation \citep{spinelli2019} or ionospheric effects \citep{Shen2020} to be accounted for by expanding the physical model to include them. The simulations using the REACH pipeline in Extended Data Fig. 3 assume the antenna beams are known exactly. As this in not possible in practice, we intend to develop this pipeline further to also fit for uncertainties in the beam model.

\subsection{\label{sec:multiantenna}Time- and antenna-dependant modelling}
The REACH data analysis pipeline is also designed to allow changes in the foregrounds and chromatic distortion due to changing observing time or from using different antennae to be exploited to model the foregrounds and 21-cm signal more accurately. Because the foreground parameters being fit for by the pipeline are a physical property of the radio sky, the spectral index, the true values of these parameters should be unchanged for different times of observation and for observations with different antennae. The same is true for the 21-cm signal parameters.

As a result of this, it is possible with this pipeline to fit many data sets from different observing times and different antennae to corresponding models simultaneously in one likelihood, with them all informing the same parameter values. The effect of doing so, as opposed to fitting a single, time-integrated data set to a single model, is shown in Extended Data Fig. 4.

Extended Data Fig. 4 top-left shows that by fitting simulated data from both a log spiral and a hexagonal dipole antenna simultaneously, the accuracy and precision of the recovered 21-cm signal are both improved. Signal accuracy is also seen to improve by fitting the data time bins jointly as separate data sets rather than integrating them. Furthermore, Extended Data Fig. 4 top-right shows that for individual antennae, fitting data time bins separately results in an increase in the optimum number of foreground parameters needed, and so a more detailed model of the foregrounds in reconstructed, relative to integrated data. However, when fitting data from both antenna simultaneously, both the time-separated and time-integrated versions require equally high detail in the foreground model.

As fitting data from multiple antenna simultaneously in particular can be seen from these results as producing improvement in both the detail of the foreground model and the accuracy of the recovered 21-cm signal, REACH is intending to deploy two antenna in order to be able to jointly fit data in this manner.

\subsection{Detection of systematic errors}

As detailed above, one of the driving principles when developing REACH has been the desire to account for and model systematics in the data analysis pipeline. For example, where experiments such as SARAS2 \citep{Singh2018a} assumed an achromatic beam, REACH is using a data driven development of the antenna and will account for chromaticity in the beam pattern directly with the foreground modelling.
	
However, regardless of how careful we are with our data analysis and calibration, the potential for unaccounted for systematics, not modelled in the Bayesian pipeline, to enter the data remains. We have seen with existing experiments that systematics can cause uncertainty in the presence of global 21-cm signals \citep{2018Natur.564E..32H, 2020MNRAS.492...22S, Bevins2020} and in some instances obscure any potential signal \citep{Price2018}. Should the data from Phase I of REACH contain any unaccounted for systematics then a quick and accurate characterisation of those systematics can lead to an identification of their cause and iterative improvements in the experimental system, calibration and/or data analysis. 

Since the REACH Bayesian pipeline assumes the presence of specific known systematics, such as chromaticity in the beam and ionospheric effects, and then corrects for these alongside the foreground modelling the presence of unaccounted for systematics has the potential to distort these corrections and consequently the foreground model. This is in particular true of the chromaticity correction which is driven by the raw data. This means that any residual systematics in the data after removing the modelled foreground and applying the distorted corrections may not be representative of the true unaccounted for systematics. To accurately identify and characterise them, should they be shown to be present, we need, therefore, to separately model the foreground and known systematics like chromaticity as accurately as possible. Particularly the known systematics have to be modelled in a way that is either independent of the raw data or that includes comprehensive modelling of each component in the data including the unaccounted for systematic.

Maximally Smooth Functions~(MSFs) have been shown to be a useful foreground modelling technique for global 21-cm experiments \citep{SathyanarayanaRao2017} and it has also been demonstrated that they can be used to accurately identify non-smooth systematics in data sets \citep{2019ApJ...880...26S, Bevins2020}. They are functions that are constrained such that
\begin{equation}
\frac{d^my}{dx^m}~\geq0~~\textnormal{or}~~  \frac{d^my}{dx^m}~\leq0,
\end{equation}
meaning that they are characteristically smooth and can act as an effective model for the smooth synchrotron and free-free foregrounds in global 21-cm experiments. The constraint also prevents the functions fitting out any non-smooth structure such as signals or systematics in the data set as can happen with an unconstrained polynomial. In the event that the Bayesian data analysis pipeline detailed above does not confidently identify a global 21-cm signal and illustrates the presence of systematics in the REACH data we will use \texttt{maxsmooth}~\citep{Bevins2020, Bevins2020b} to fit MSFs to the calibrated data and help identify unaccounted for systematics.

In order to do this, we will need to apply separate data independent corrections for the expected chromaticity from the antenna beams and for other known systematics to the data. This will ensure that our MSF best represents the foreground and any remaining residuals after modelling correspond to the unaccounted for systematics.

We will attempt to physically model the structure left in the residuals after modelling known systematics and the smooth foreground. For example, any periodicity to the residuals could be linked to the reflections in cables or ground emission and the depth of discontinuities in the soil surrounding the antenna. By subsequently wrapping \textsc{maxsmooth} inside a Bayesian nested sampling loop with and without various model components, including our physical model for the unaccounted for systematic and various signal profiles, we can use the evidence of the different fits to determine whether the data favours the presence over the absence of our physical model or not. This can help provide confidence in our characterisation of the unaccounted for systematic or indicate that new physics is needed to describe the structure of the signals in the REACH data.

Assuming a correct characterisation and functional description of any unaccounted for systematics, we can confidently identify a cause which can then be mitigated in future iterations of the experiment. We can also use the functional description in unison with the Bayesian pipeline to fit for the unaccounted for but subsequently characterised systematics, chromaticity correction, additional known systematics and foreground using nested sampling.

We provide a demonstration of the use of MSFs on EDGES low band data in the \textit{supplementary information document - Figure 7.}

\subsection{Systematic signals in the data pipeline}

Once a systematic error signal has been detected (for example by running \texttt{maxsmooth} on the data) and identified, either it is corrected in hardware (eg. via hardware specific modifications/upgrades) or a model of the signal is included in the data analysis pipeline. Two possibilities exist in the later case, either the model is parametric and the parameters can be jointly fitted together with the foreground and the cosmological signals or the the model is fixed. 

Furthermore, in order to understand the ability of the data pipeline to cope with specific systematic signals expected in the REACH experiment, a series of simulated data analysis including such signals are performed. In Extended Data Fig. 4 bottom-left an example of such simulation is shown. In this simulation, a non-parametric (fixed) model of the antenna beam in which the presence of the finite metallic ground plane (20x20 m) below the antenna is used to both generate the data and for the model of the data pipeline. We note that this is not the ideal case, since ideally a parametric-model of the beam capable of representing the effects of the ground plane is available and can be used. However, this simulation allows us to understand if the chromaticity introduced by a specific feature of the instrument design would result in the Bayesian analysis failing in the same way that certain antenna designs can do that. In Fig. Extended Data Fig. 4 bottom right, a simulated analysis where a specific systematic expected from the REACH design is both introduced in the data as well as modelled with a parametric model and fitted during the data analysis, is shown. This simulation is for a systematic signal corresponding to uncalibrated reflections in the 6 m coaxial cable used to connect the conical log-spiral antenna to the receiver box. We note that for the dipole antenna this effect is a lot less worrying since the cable is much shorter (<0.5 m) and therefore any reflections produce data at scales of less concern for the detection of the global 21-cm signal. This simulation demonstrates that if an adequate parametric model is available, it is potentially possible to avoid the detrimental effect of a systematic signal and achieve a detection. Since the Bayesian evidence of these fits is available to us, it can be determined if the presence of the systematic signal is favoured or not. We note that currently we are developing parametric models for the antenna beam.

While these are scenarios where we have identified a potential systematic signal, the results help to demonstrate our ability to potentially make a detection even in the presence of such systematic signals. More information on this can be found in \citep[][]{ScheutwinkelB}.

\subsection{Cosmological models}

The 21-cm signal simulations considered in this paper assume a standard astrophysical scenario (with the CMB as background radiation and astrophysical channels for cooling and heating of the gas). 

The simulation methodology is semi-numerical and initially generates  cosmological boxes of a few hundreds comoving Mpc in which large scale structure is evolved. The large size of the cosmological boxes is necessary to account for the non-linear dependence of the global 21-cm signal on non-local astrophysical phenomena. Star formation is included at sub-grid level, as the simulations are aimed at large-scale 21-cm signals. The simulations are  initialized with cubes of density, temperature  and relative velocity between dark matter and baryons \citep{tseliakhovich10}. The density and velocity fields are evolved using linear perturbation theory. The number of dark matter halos in each resolution element (cell of 3$^3$ comoving Mpc$^3$) is determined based on the values of the local density and relative velocity and is derived at each redshift using a modified Press-Schechter model \citep{press74, sheth99, barkana04}. 
Gas in dark matter halos above the star formation threshold, parameterized by minimum circular velocity $V_\mathrm{c}$ measured in km s$^{-1}$, is converted into stars with efficiency of $f_\mathrm{*}$. 

Given a population of galaxies, radiation fields that interact with the intergalactic medium (IGM), and ultimately determine the strength of the 21-cm signal, are calculated as follows. Cosmic heating has several contributions including X-ray binaries (XRBs) \citep{fragos13}, which are the dominant source of heating in most astrophysical cases, Ly-$\alpha$ \citep{chuzhoy07, 10.1093/mnras/staa3811} and the CMB \citep{venumadhav18}. The heating rate by high redshift XRBs depends on the properties of these sources, most importantly on their efficiency $f_X$ (defined as the ratio of the bolometric X-ray luminosity to star formation rate) and spectral energy distribution (SED) \citep{fialkov14a, Pacucci:2014, fialkov14, cohen18}, which is modelled as a power-law with a slope $\alpha$ and cutoff frequency $\nu_{\rm min}$. Reionization of the IGM is implemented using the excursion set formalism \citep{Furlanetto:2004} and is subject to the photoheating feedback \citep{cohen16}. The process of reionization is parameterized by the CMB optical depth $\tau$ (itself, a function of the ionizing efficiency of star forming galaxies, $\zeta$) and mean free path of ionizing photons R$_{\rm mfp}$ (measured in comoving Mpc).

\subsection{Additional science outputs}

Below we present a list of additional science outputs for the REACH experiment.

\begin{itemize}
    \item In the case of non-detection, upper limits can be derived on the strength of the absorption feature, which will allow us to put constraints on the astrophysical properties of sources. In particular, if no signal is detected by REACH, that would either imply that the signal is either (i) below the detection threshold of the instrument, which would require X-ray sources, present prior to efficient WF-coupling of the spin and kinetic temperatures of the gas, to be very luminous, or (ii) outside the REACH band, which would constrain star formation to happen very late and in very massive dark matter halos. 

    \item As illustrated in Fig. \ref{Fig:science_forecasts2}, a high signal-to-noise detection of the 21-cm signal with REACH will enable us to place competitive independent constraints on $\tau_\mathrm{CMB}$. In cosmological inference from CMB data, the amplitude of the CMB angular power spectrum is proportional to the degenerate product $A_{s}e^{-2\tau_\mathrm{CMB}}$, where $A_{s}$ is the amplitude of primordial density fluctuations. Independent constraints on $\tau_\mathrm{CMB}$ break this degeneracy; thus, joint analysis of REACH and Planck data will enable more stringent limits on $A_{s}$ to be obtained. This, in turn will enable us to bring into sharper relief current tensions between the $A_{s}$ inferred using the CMB and directly measured by large-scale structure probes (e.g. Sunyaev–Zeldovich cluster counts, galaxy lensing, Baryon Acoustic Oscillations measurements) on smaller scales (e.g. \cite{2015PhRvD..91j3508B}). Furthermore, this also would potentially provide science output in inflationary physics, where $\tau_\mathrm{CMB}$ is the only way to get truly closer to cosmic variance limited science \citep{2018JCAP...04..016F}. We note that the degeneracy between $A_{s}$ and $\tau_\mathrm{CMB}$ can also be reduced using estimates of $\tau_\mathrm{CMB}$ from large angular scale modes of CMB polarization power spectra or using CMB lensing if one assumes no departures from standard $\Lambda\mathrm{CDM}$ cosmology.
    \item REACH as a gravitational wave detector. It is being postulated \cite{PhysRevLett.126.021104} that gravitational waves can be converted into photons in the presence of magnetic fields. This in turn can lead to distortion of the CMB radiation measurable at MHz frequencies with radio telescopes such as REACH. Establishing upper bounds on Gravitational Waves is therefore a potential high impact additional science outcome of REACH. 
    \item Better low frequency sky models constrained by REACH data. The intrinsic joint analysis performed by REACH means that simultaneously with cosmological models, REACH will produce accurate measurements of foreground model parameters. Therefore, despite its lack of high resolution on the sky, we anticipate significant science output from analyzing these foreground model fits. Examples include establishing accurate measurements of the spectral index of different large regions of the sky or potential measurements of absolute power of the diffuse emission from the foregrounds. 
    \item Ionosphere and space weather science. REACH is essentially an extremely accurate all sky (averaged sky) monitor. We expect to be able to use REACH to improve our understanding of the temporal and spectral fluctuations of the large spatial scales of the ionosphere as well as of space weather phenomena (eg. solar activity through its interaction with the ionosphere). 
    \item Serendipitous science. With an extremely accurate and precise radiometer of electromagnetic waves from the sky at long wavelengths (aiming at being the most accurate and precise at frequencies 50 - 170 MHz), we anticipate the possibility of unexpected discoveries, as is typically the case when technology is pushed to its limits. 
\end{itemize}

\subsection{Data availability}

Upon detection or significant scientific result our data will be made publicly available on Zenodo.

\subsection{Code availability}

Upon detection or significant scientific result our code will be made publicly available on GitHub.

The \texttt{maxsmooth} code can be found online at https://github.com/htjb/maxsmooth. The \texttt{globalemu} code can be found online at https://github.com/htjb/globalemu.

\section*{Corresponding author}

Correspondence and requests for materials should be addressed to Eloy de Lera Acedo at eloy@mrao.cam.ac.uk.
\\
\begin{acknowledgments}
The REACH collaboration acknowledges The Kavli Institute for Cosmology in Cambridge (www.kicc.cam.ac.uk), Stellenbosch University (www.sun.ac.za), the National Research Foundation of South Africa (www.nrf.ac.za), and the Cambridge-Africa ALBORADA Research Fund (www.cambridge-africa.cam.ac.uk/initiatives/the-alborada-research-fund/) for their financial support of the project. E. de Lera Acedo wishes to acknowledge the support of the Science and Technology Facilities Council (STFC) through grant number ST/V004425/1 (Ernest Rutherford Fellowship). G.~Bernardi and M.~Spinelli acknowledge support from the Ministero degli Affari Esteri della Cooperazione Internazionale - Direzione Generale per la Promozione del Sistema Paese Progetto di Grande Rilevanza ZA18GR02 and the National Research Foundation of South Africa (Grant Number 113121) as part of the ISARP RADIOSKY2020 Joint Research Scheme. The research of D.I.L. de Villiers and O. Smirnov is supported by the South African Research Chairs Initiative of the Department of Science and Technology and National Research Foundation. M.~Spinelli acknowledge funding from the INAF PRIN-SKA 2017 project 1.05.01.88.04 (FORECaST). This work is based on the research supported in part by the National Research Foundation of South Africa (Grant Number 75322). H. T. J. Bevins acknowledges the support of STFC through grant number ST/T505997/1.
\end{acknowledgments}

\section*{Author contributions statement}

EdLA is the PI and initiator of REACH, led the paper write-up coordination and the introduction, experimental approach, architecture, high-level system metrics, data pipeline and instrument models sections; DILdV is the co-PI of REACH and led the antenna and site/RFI sections; NRG led the receiver calibrator section; WH led the Bayesian data analysis section; AF co-led the science prospects section, led the cosmological models section and contributed to the introduction section; AM led the digital back-end section; DA led the EDGES data re-analysis, data analysis driven antenna selection, foreground models and chromaticity correction and time- and antenna-dependant sections; HTJB led the detection of systematic errors section; RC contributed significantly to the digital back-end section; JC contributed significantly to the antenna section; ATJ contributed significantly to the site/RFI section; ILVR led the Bayesian receiver calibration section; PHS co-led the science prospects section; KHS led the Systematic signals in the data pipeline section. The rest of the authors (PA, GB, SC, JCa, WC, JAE, TGJ, QG, RH, GK, RM, PDM, SM, JRP, EP, AS, ES, OS, MS and KZA) have contributed to writing different sections and to review the manuscript.
\\
\section*{Competing Interests Statement}

The authors declare no competing interests.
\\

\section*{Tables}

\begin{table*}
\centering
\begin{tabular}{c c c m{2cm} m{2.5cm} m{2.6cm} c}
\hline 
\textbf{Experiment} & \textbf{Bandwidth} & \textbf{Freq. band /MHz} & \textbf{Simultaneous observations} & \textbf{Antenna type} & \textbf{Receiver calibrator and spectrometer} & \textbf{Full Bayesian analysis}\\
\hline
\hline EDGES &2:1& 50-100 (LB), 100-200 (HB)  & No & Blade dipole & Lab measurements, auto-correlation & No\\
\hline SARAS& 5:1&40-200 (SARAS2)&No&Monopole&Lab measurements, cross-correlation &No\\
\hline LEDA&2.125:1&30-85&No&Crossed drooping dipole&Lab measurements, auto-correlation  &No\\
\hline PRIZM&2:1&50-90, 70-140&No&Crossed-dipole (four-square)&Lab measurements, auto-correlation  &No\\
\hline SCI-H I&3.21:1&40-130&No&HIbiscus&Lab measurements, auto-correlation &No\\
\hline BIGHORNS&2.85:1&70-200&No&Conical log-spiral&Lab measurements, auto-correlation  &No\\
\hline \textbf{REACH}&\textbf{3.4:1}&\textbf{50-170}&\textbf{2 antennas}&\textbf{Hexagonal dipole, Conical log-spiral}&\textbf{In-field measure., auto-correlation }&\textbf{Yes}\\

\hline
\end{tabular}
\caption{\label{tab:Comparison}Comparison of main experimental features of existing global experiments with REACH. REACH, to our best knowledge, is the only ground-based 21-cm global experiment featuring a full joint Bayesian model fitting including all signal components.}
\label{comp}
\end{table*}

\section*{Figure Legends/Captions}

\bibliographystyle{unsrt}
\bibliography{reach}

\begin{figure*}
\centering
\includegraphics[width=\linewidth]{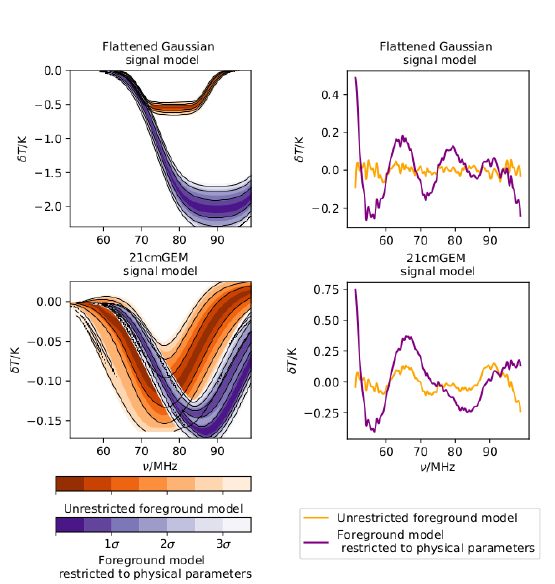}
\textbf{EXTENDED DATA FIGURE 1} Re-analysis of the publicly available EDGES data restricting the foregrounds to physical parameters (purple) and with unrestricted foreground parameters (orange) for a flatten Gaussian EDGES-style signal model (top-left) and a \texttt{21cmGEM} standard physical model from \citep{cohen19} (bottom-left). On the right column we show the corresponding residuals after subtraction of the posterior average fitted foreground and signal models.
\label{fig:EDGES}
\end{figure*}

\begin{figure*}
\centering
\includegraphics[width=\linewidth]{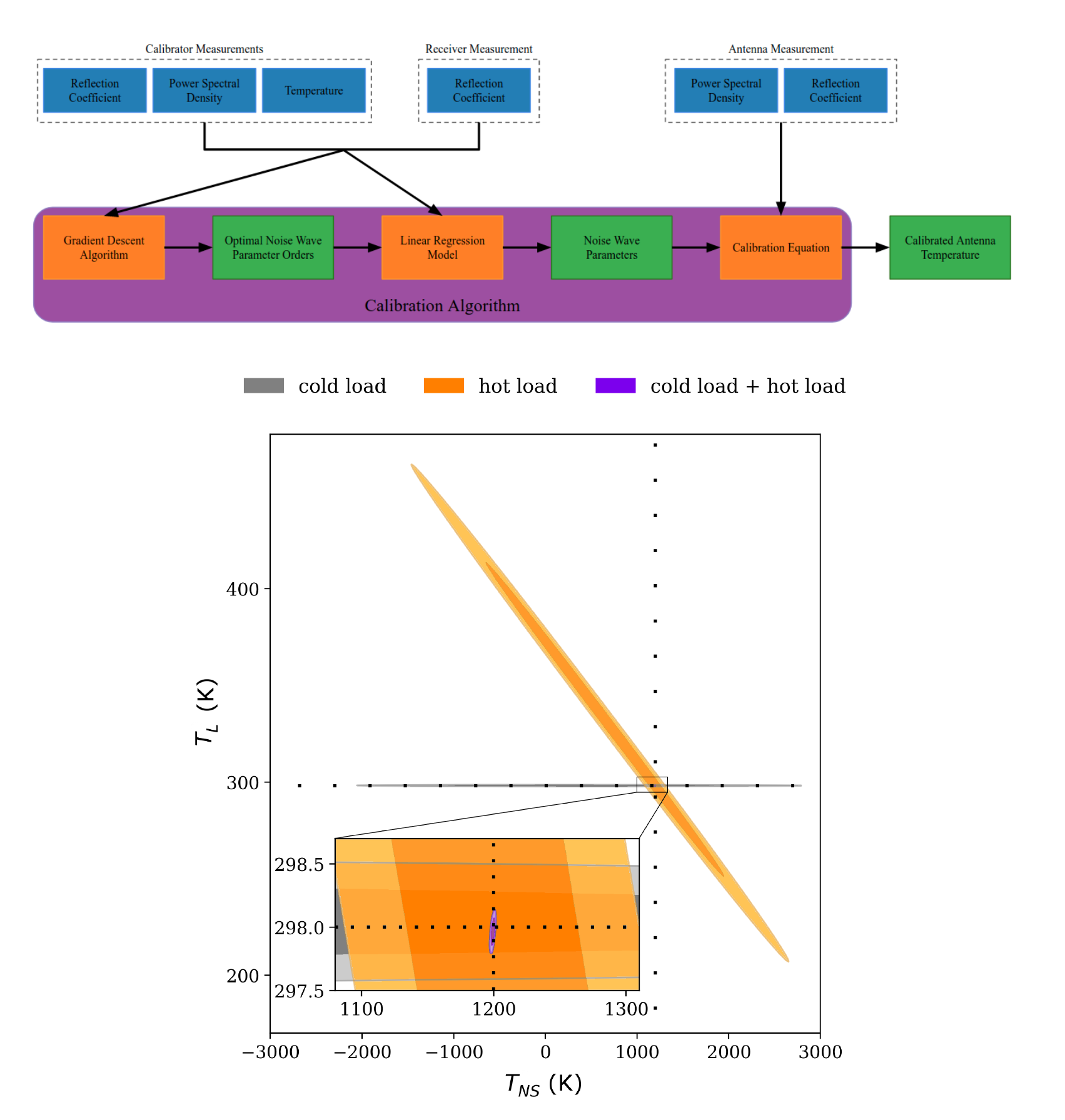}
\textbf{EXTENDED DATA FIGURE 2} Top: Outline of the calibration algorithm. Blue blocks represent data to be taken, red blocks represent calculations and green blocks represent calculation outputs. Bottom: Plot showing the joint posteriors for two noise wave parameters used for calibration of the receiver; $T_{\mathrm{L}}$ and $T_{\mathrm{NS}}$. Posteriors are derived using a single room-temperature `cold' load as a calibrator, a single `hot' load heated to \mbox{373 K} and both loads used in tandem shown in grey, red and blue respectively. The black cross hairs mark the known values of the calibration parameters. A zoom-in of the posterior intersection is provided to illustrate the constraint on parameter values attributed to the correlation between parameters that is considered by our algorithm when deriving the blue, dual-load posterior.
\label{fig:RXcalibration}    
\end{figure*}

\begin{figure*}
    \centering
    \includegraphics[width=0.9\linewidth]{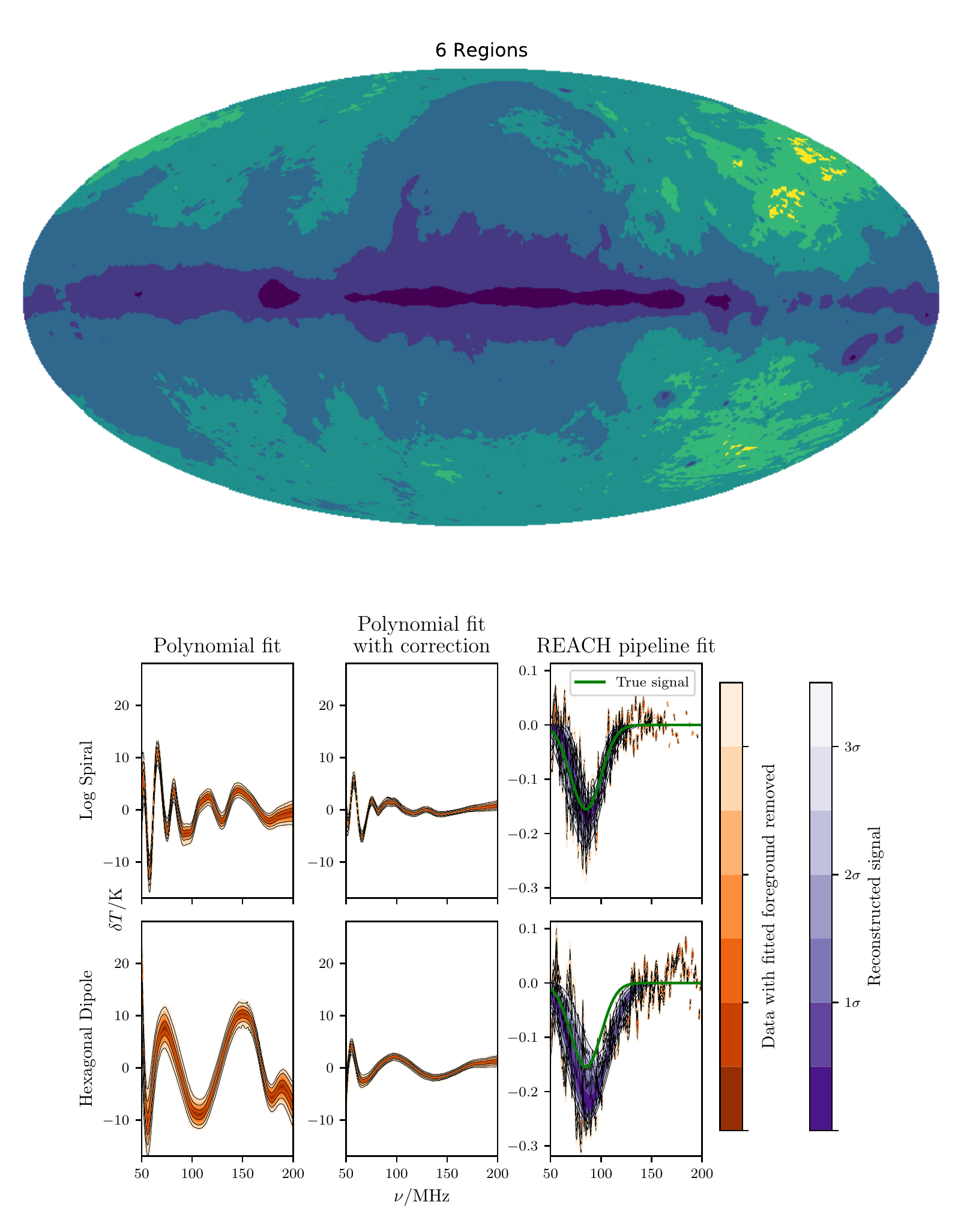}
     \textbf{EXTENDED DATA FIGURE 3} Top: Plot showing the subdivision of the sky in galactic coordinates into a number of regions $N=6$ of similar spectral index. Bottom: Plot comparing the residuals from fitting simulated 21-cm data. The plots shows the results of fitting data with a 5th order log-polynomial model (left), fitting data corrected by (\ref{eq:FoMchrom}) with a 5th order log-polynomial model (centre) and fitting the data with the REACH pipeline, using $N=9$ (right). The residuals after subtraction of the foreground models are shown in red. The signal model and true signal inserted into the simulated data, are shown in blue and green respectively, where visible. These results are simulated using a conical log-spiral antenna and a hexagonal dipole antenna.
     \label{fig:regions}
\end{figure*}

\begin{figure*}
    \centering
    \includegraphics[width=0.7\textwidth, angle=270]{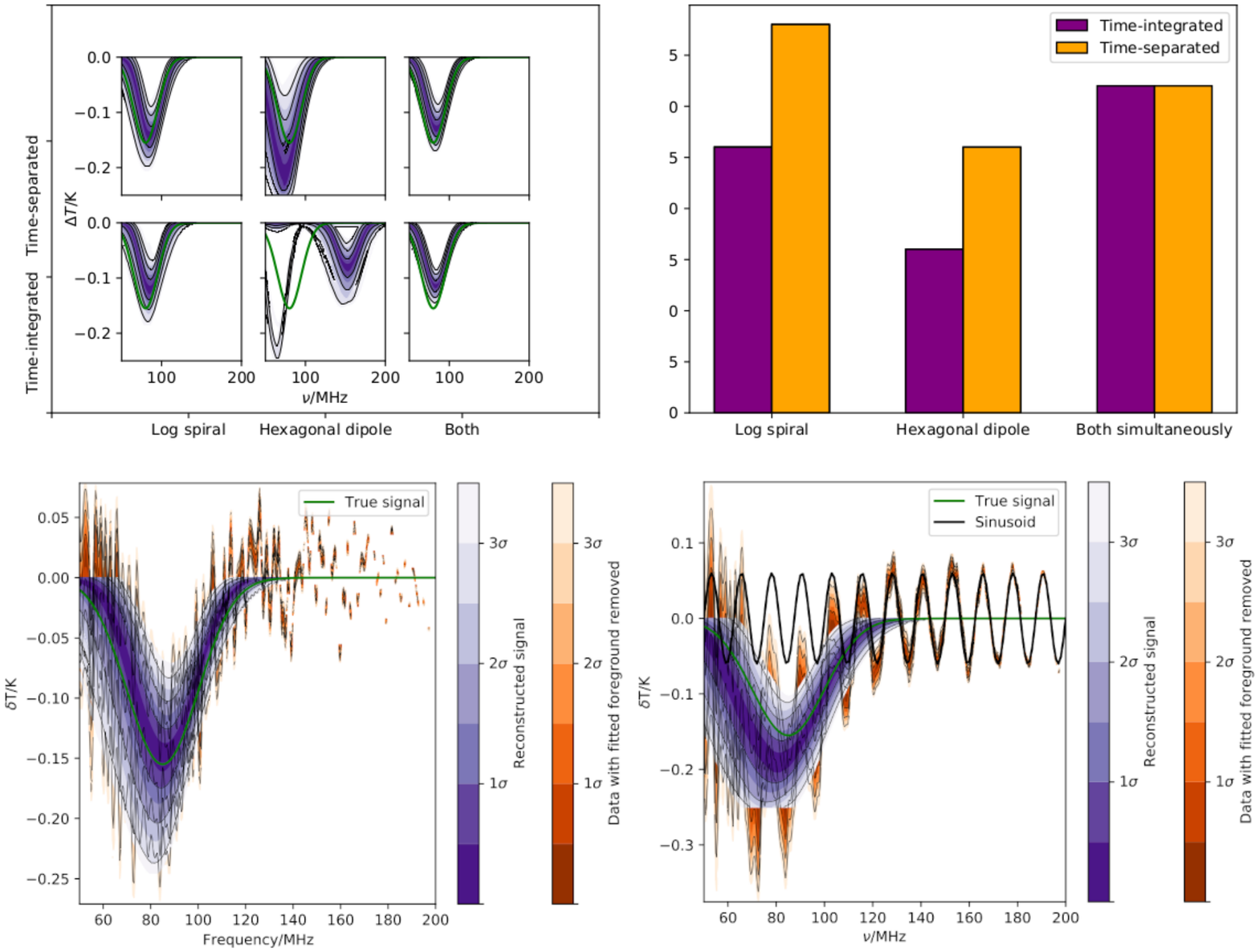}
    
    \textbf{EXTENDED DATA FIGURE 4} Top-left: Plot of the recovered 21-cm signal in purple, compared to the true inserted 21-cm signal in green, for simulated data sets of a log spiral and hexagonal dipole antenna. Each data set consisted of three time bins, 20 minutes apart. The lower plots show the results of fitting an integration of the three bins to a single foreground model and the upper plots show the results of fitting the separate bins jointly to corresponding models in a single fit. The rightmost plots show the results of fitting the data sets from both antenna simultaneously in the same fit. Top-right: Plot of the optimum numbers of foreground regions, determined using the Bayesian evidence, for the model fits shown in the top-left plot. Bottom-left: Plot showing a run of the pipeline where the antenna model included the presence of the finite 20x20 m metallic ground plane underneath the spiral antenna. This plot shows that the chromaticity introduced by reflections at the edge of the REACH ground plane, if properly modelled, would not severely affect the ability of the pipeline to recover the cosmological signal. Bottom-right: Plot showing the result of running the data pipeline when a sinusoidal systematic arising from the presence of the 6 m cable connecting the spiral antenna feed point to the receiver has been introduced in the data. The additive systematic signal is shown as a black-solid line in this plot. In the simulated analysis we included a sinusoidal model to fit for this systematic signal simultaneously with the foregrounds and the cosmological signal. This result shows that a detection of the true signal could be achieved in this case.
    \label{fig:multi_antenna_sys}
    
\end{figure*}

\end{document}